%% file: 0ieee.tex
\documentclass[conference,compsoc]{IEEEtran}

\input{00macros}

\includeversion{arxiv}
\excludeversion{submit}

\begin{document}

\title{SoK: Unintended Interactions among Machine Learning Defenses and Risks}

\author{
    \IEEEauthorblockN{Vasisht Duddu\IEEEauthorrefmark{1}, Sebastian Szyller\IEEEauthorrefmark{2}, N. Asokan\IEEEauthorrefmark{1}\IEEEauthorrefmark{3}}
    \IEEEauthorblockA{\IEEEauthorrefmark{1}University of Waterloo, \IEEEauthorrefmark{2}Intel Labs, \IEEEauthorrefmark{3}Aalto University}
    \IEEEauthorblockA{vasisht.duddu@uwaterloo.ca, contact@sebszyller.com, asokan@acm.org}
}

\maketitle

\begin{abstract}
\input{01abstract}
\end{abstract}

\input{1introduction}
\input{2background}
\input{3framework}
\input{4.1protectionmech}
\input{4.2risks}
\input{4.3interactions}

\input{4.4guideline}
\input{5conjectures}
\input{6related}
\input{7discussion}

\section*{Acknowledgements}
This work is supported in part by Intel (Private AI consortium) and the Government of Ontario. Views expressed in the paper are those of the authors and do not necessarily reflect the position of the funding agencies.

% \newpage
{\footnotesize
\bibliographystyle{IEEEtranS}
\bibliography{paperS}
}

\input{8appendix}

\end{document}

%% file: 00macros.tex
\usepackage{multirow}
\usepackage{colortbl}
\usepackage{hhline}
\usepackage{subfigure}
\usepackage{pifont}
\usepackage{rotating}
\usepackage{cite}
\usepackage{amsmath,amssymb,amsthm}
\usepackage{tikz}
\usepackage[colorlinks=true]{hyperref}
\hypersetup{citecolor=black,linkcolor=black,urlcolor=black}
\usetikzlibrary{shapes,arrows,backgrounds, positioning, fit}
\usepackage{verbatim}
\usepackage[inline]{enumitem}

\usepackage[framemethod=TikZ]{mdframed}
\usepackage{xspace}
\usepackage[most]{tcolorbox}
\usepackage{booktabs}
\usepackage[T1]{fontenc}
\usepackage{versions}
\usepackage{xcolor}
\definecolor{ultramarine}{RGB}{250,175,92}
\newcommand\change[1]{{\color{purple}{#1}\xspace}}

\usepackage{wasysym}
\newcommand{\conjmark}{\varhexstar\xspace}
\newcommand{\cmark}{\CIRCLE\xspace}
\newcommand{\xmark}{\Circle\xspace}
\newcommand{\smark}{\astrosun\xspace}

\newcommand{\redmark}{{\color{red!60}\cmark}}
\newcommand{\graymark}{{\color{gray!60}\cmark}}
\newcommand{\greenmark}{{\color{green!60}\cmark}}
\newcommand{\race}{\texttt{Race}\xspace}
\newcommand{\sex}{\texttt{Sex}\xspace}
\newcommand{\newtext}[1]{\textcolor{black}{#1}}

\newcommand{\greensquare}{\aquarius\xspace}
\newcommand{\redsquare}{\DOWNarrow\xspace}

\newcommand{\factor}{\texttt{<f>}\xspace}
\newcommand{\commonfactor}{\texttt{<f$_{c}$>}\xspace}
\newcommand{\protmech}{\texttt{<d>}\xspace}
\newcommand{\risk}{\texttt{<r>}\xspace}

\newcommand{\attmodel}{f_{att}\xspace}
\newcommand{\dtrain}{\mathcal{D}_{tr}\xspace}
\newcommand{\dtest}{\mathcal{D}_{te}\xspace}
\newcommand{\dtrainadv}{\mathcal{D}^{adv}_{tr}\xspace}
\newcommand{\dtestadv}{\mathcal{D}^{adv}_{te}\xspace}

\newcommand{\adv}{\protect{$\mathcal{A}dv$}\xspace}
\newcommand{\model}{f_{\theta}\xspace}
\newcommand{\robustmodel}{f^{rob}_{\theta}\xspace}
\newcommand{\privatemodel}{f^{dp}_{\theta}\xspace}
\newcommand{\fairmodel}{f^{fair}_{\theta}\xspace}
\newcommand{\fairmodelshadow}{\bar{f}^{fair}_{\theta}\xspace}
\newcommand{\noOutlierModel}{f^{outrem}_{\theta}\xspace}
\newcommand{\watermarkModel}{f^{wm}_{\theta}\xspace}
\newcommand{\derivedmodel}{f^{der}_{\theta}\xspace}

\newcommand{\loss}{\mathit{l}}
\newcommand{\Loss}{\mathrm{L}}
\newcommand{\expected}[2]{\underset{#2}{\mathbb{E}}\Large[#1\Large]}

\newcolumntype{P}[1]{>{\centering\arraybackslash}p{#1}}
\newcommand{\cmmnt}[1]{}

%% file: 01abstract.tex
Machine learning (ML) models cannot neglect risks to security, privacy, and fairness. Several defenses have been proposed to mitigate such risks. When a defense is effective in mitigating one risk, it may correspond to increased or decreased susceptibility to \emph{other risks}. Existing research lacks an effective framework to recognize and explain these \textit{unintended interactions}. 
We present such a framework, based on the conjecture that \emph{overfitting} and \emph{memorization} underlie unintended interactions. We survey existing literature on unintended interactions, accommodating them within our framework. We use our framework to conjecture two previously unexplored interactions, and empirically validate them.
\begin{arxiv}
\footnote{IEEE Symposium on Security and Privacy (S\&P) 2024}
\end{arxiv}

%% file: 1introduction.tex
\section{Introduction}\label{introduction}

Potential \textit{risks} to security, privacy, and fairness of machine learning (ML) models have led to various proposed \textit{defenses} ~\cite{robustSoK,SoKMLPrivSec,certifiedRobustness2020sok,poisonSurvey,lukas2022sok,privacySurvey,fairSurvey,transparencySurvey}.
When a defense is effective against a specific risk, it may correspond to an increased or decreased \emph{susceptibility to other risks}~\cite{surveyTradeoff,shokriSurvey}. For instance, adversarial training for increasing robustness also increases susceptibility to \emph{membership inference}~\cite{MIARobustness,Hayes2020TradeoffsBM}. We refer to these as \textit{unintended interactions} between defenses and risks.

Foreseeing such unintended interactions is challenging. A unified framework enumerating different defenses and risks can help (a) researchers identify unexplored interactions and design algorithms with better trade-offs, and (b) practitioners to account for such interactions before deployment. However, prior works are limited to a specific risk (e.g., membership inference~\cite{shokriSurvey}), defense (e.g., fairness with privacy risks~\cite{fedtradeoff}) or interaction (e.g., explanations with security risks~\cite{sokExpl}) and do not systematically study the underlying causes (e.g., Gittens et al.~\cite{surveyTradeoff}). A comprehensive framework spanning multiple defenses and risks, to systematically identify potential unintended interactions is currently absent.

We aim to fill this gap by systematically examining various unintended interactions spanning multiple defenses and risks. 
We conjecture that \emph{overfitting} and \emph{memorization} of training data are the potential causes underlying these unintended interactions. 
An effective defense may induce, reduce or rely on overfitting or memorization which in turn, impacts the model's susceptibility to other risks.
We identify different factors like the characteristics of the training dataset, objective function, and the model, that collectively influence a model's propensity to overfit or memorize. %(e.g., size, number of attributes, distribution); (e.g., curvature smoothness, distinguishability of model observables, and distance to decision boundaries); (e.g., capacity)
These factors help us get a nuanced understanding of the susceptibility to different risks when a defense is in effect. 
We claim the following main contributions:
\begin{enumerate}[leftmargin=*]
\item the \emph{first systematic framework} for understanding unintended interactions, by their underlying causes and factors that influence them. 
(Section~\ref{sec:framework})
\item a literature \emph{survey} to identify different unintended interactions, situating them within our framework, and a guideline to use our framework to conjecture unintended interactions. (Section~\ref{sec:interactions})
\item identifying \emph{unexplored unintended interactions} for future research, using our guideline to \emph{conjecture two such interactions}, and \emph{empirically validating} them. (Section~\ref{sec:conjectures})
\end{enumerate}

%% file: 2background.tex
\section{Background}\label{sec:background}

We introduce common ML concepts (Section~\ref{sec:mlback}), different risks to ML (Section~\ref{sec:backrisk}) and corresponding defenses proposed in the literature (Section~\ref{sec:backprotection}).

\subsection{ML Classifiers}\label{sec:mlback}

Let $X$ be the space of input attributes, $Y$, the corresponding labels, and $P(X,Y)$ the underlying probability distribution of all data points in the universe $X \times Y$. A training dataset $\dtrain = \{x_i,y_i\}_{i=1}^n$ is sampled from the underlying distribution, i.e., $\dtrain \sim P(X,Y)$. The $i^{th}$ data record in $\dtrain$ contains input attributes $x_i \in X$ and the classification labels $y_i \in Y$. An ML classification model $\model$ is a parameterized function mapping: $\model: X \rightarrow Y$ where $\theta$ represents $\model$'s parameters. $\model$'s capacity is indicated by the number of parameters which varies with different hyperparameters such as the number, size, and types of different layers.

\noindent\textbf{Training.} For each data record $(x, y) \in \dtrain$, we compute the loss $\loss(\model(x), y)$ using the difference in model's prediction $\model(x)$ and the true label $y$. $\theta$ is then iteratively updated to minimize the cost function $\mathcal{C}$ defined as: $\min\limits_{\theta} \, L(\theta) + \lambda \, R(\theta)$ where $\Loss(\theta) = \expected{\loss(\model(x),y)}{(x,y) \sim \dtrain}$. Here, $R(\theta)$ is the regularization function which is used to restrict the range of $\theta$ to reduce overfitting to $\dtrain$. $\lambda$ is the regularization hyperparameter which controls the extent of regularization. The loss function is minimized using gradient descent algorithm (e.g, Adam) where $\theta$ is updated as $\theta := \theta - \frac{\partial \mathcal{C}}{\partial \theta}$ where $\frac{\partial \mathcal{C}}{\partial \theta}$ is the gradient of the cost with respect to $\theta$. Training stops when algorithm converges to one of $\mathcal{C}$'s local minima. 
% We refer to the hyperparameters used during training such as the learning rate, and number of epochs, choice of optimization algorithm, objective function and regularization as \textit{training configuration}.

\noindent\textbf{Inference.} Once $\model$ is trained, clients can query $\model$ with their inputs $x$ and obtain corresponding predictions $\model(x)$ (blackbox setting as in ML as a service) or intermediate layer activation $f^k_{\theta}(x)$ and $\theta$ (whitebox setting as in federated learning). We refer to $\model(x)$, $f^k_{\theta}(x)$, $\frac{\partial \mathcal{C}}{\partial \theta}$ corresponding to an input, and $\theta$ as \textit{observables}. We evaluate the performance of $\model$ using accuracy on an unseen test dataset ($\dtest$). 

\subsection{Risks to ML}\label{sec:backrisk}

We identify three categories of risk:
\begin{itemize}[leftmargin=*,wide,  labelindent=0pt]
\item \underline{\textbf{Security}} includes \textit{evasion}, \textit{poisoning}, and \emph{theft}:
\begin{enumerate}[label=\textbf{R\arabic*},leftmargin=*,wide,  labelindent=0pt]
\item\label{r1} \textbf{Evasion} forces $\model$ to misclassify on specific inputs with carefully crafted perturbations~\cite{goodfellow2014explaining,robustSoK}. Given an input $x$, an adversary \adv generates an adversarial example $x_{adv}$ by adding a perturbation $\delta$ (bounded by $\epsilon_{rob}$), i.e., $x_{adv} = x + \delta$ which maximizes $\model$'s loss: $\max_{\|\delta\| \leq \epsilon_{rob}}\ell(\model(x+\delta), y)$. %This perturbation is bounded within an $l_p$ norm, i.e., $||\delta||_p<\epsilon_{rob}$ where $\epsilon_{rob}$ is the perturbation budget.

\item\label{r2} \textbf{Poisoning} modifies $\model$'s decision boundary by adding malicious data records to $\dtrain$ (referred as \textit{poisons}). This is done to force $\model$ to misclassify specific inputs, or make $\model$'s accuracy worse. Backdoor attacks~\cite{backdoor} are a specific case when \adv trains $\model$ with some inputs with a ``trigger'' corresponding to an incorrect label. $\model$ maps any input with the trigger to the incorrect label but behaves normally on all data records. 

\item\label{r3} \textbf{Unauthorized model ownership} is when \adv obtains a derived model ($\derivedmodel$) mimicking $\model$'s functionality. %These attacks are categorized as blackbox and whitebox based on how \adv obtains $\derivedmodel$. 
It may be an identical copy of $\model$ %, e.g. by breaking into a device 
(\textit{whitebox model theft}), or %\textit{Blackbox attack} creates a copy of $\model$ 
derived by querying $\model$ and using the outputs to train $\derivedmodel$ (\textit{blackbox model extraction)})\cite{orekondy2019knockoff,correia2018copycat,stealml}. %Having an exact or a near-copy of $\model$ violates its confidentiality.
There are two types of \adv: (a) \emph{malicious suspects} perform model theft but try to evade ownership verification by modifying $\derivedmodel$ to look different from $\model$, and (b) \emph{malicious accusers} falsely accuse an independently trained model as being $\derivedmodel$~\cite{liu2023false}.
\end{enumerate}

\item \underline{\textbf{Privacy}} covers \textit{inference attacks} and \textit{data reconstruction}:

\begin{enumerate}[label=\textbf{P\arabic*},leftmargin=*,leftmargin=*,wide,  labelindent=0pt]
\item \label{p1} \textbf{Membership inference} exploits the difference in $\model$'s behaviour on data records in inside and outside $\dtrain$ to infer the membership status of a specific data record~\cite{ye2022enhanced,carlini2021membership}. 
\begin{arxiv}
This poses a privacy risk when $\dtrain$ is privacy-sensitive, such as for individuals on a dating app or with health conditions.
\end{arxiv}
\item \label{p2} \textbf{Data reconstruction} recovers entire data records in $\dtrain$ using observables~\cite{modelinvccs,genDataReconstruction,dlg,gradinvert,gradinvert2}. Furthermore, generative models can be attacked to directly output memorized data records in $\dtrain$~\cite{extractlm,secretSharer,memorizationGenerative}.

\item\label{p3} \textbf{Attribute inference} infers values of specific input attributes by exploiting distinguishability in observables. Specifically, observables are distinguishable for different values of the input attribute. This attack violates attribute privacy when the inferred attribute is sensitive~\cite{malekzadeh2021honestbutcurious,bertinoModelInv,Song2020Overlearning,fredrikson2,yeom,jayaraman2022attribute}. 
\begin{arxiv}
For instance, individuals may prefer to not disclose their race to avoid potential discrimination. % which may influence the outcome. 
\end{arxiv}

\item\label{p4} \textbf{Distribution inference} (aka property inference) infers properties of the distribution from which $\dtrain$ was sampled~\cite{propinf,propinf2,propinf3,propinf4,propinf5}. 
\begin{arxiv}
This is a privacy and confidentiality risk if the properties inferred as sensitive (e.g., potential redlining by identifying locations with minority subgroups using census information). 
\end{arxiv}
\adv exploits the difference in model behaviour when trained on datasets with different properties. 

\end{enumerate}

\item \underline{\textbf{Fairness}} includes $\model$'s discriminatory behaviour for different sensitive subgroups.

\begin{enumerate}[label=\textbf{F},leftmargin=*,leftmargin=*,wide,  labelindent=0pt]
\item \textbf{Discriminatory behaviour}\label{u} is observed when $\model$'s behaviour varies for different sensitive subgroups (e.g., ethnicity or gender) with respect to different metrics such as accuracy, false positive/negative rates. This arises from bias in $\dtrain$ or the algorithm~\cite{fairSurvey,fairSurvey2}. 
Identifying such biases is challenging as $\model$'s behaviour is \textit{incomprehensible}, i.e., difficult to explain why $\model$ made a particular prediction.
\end{enumerate}
\end{itemize}

\subsection{Defenses for ML}\label{sec:backprotection}

We identify defenses for the same categories, beginning with security (\ref{rd1}, \ref{rd2}, \ref{rd3}, and \ref{rd4}).

\begin{enumerate}[label=\textbf{RD\arabic*},leftmargin=*,leftmargin=*,wide,  labelindent=0pt]
\item\label{rd1}\textbf{Adversarial training} safeguards against \ref{r1} by training $\model$ with adversarial examples~\cite{robustSoK,certifiedRobustness2020sok}. For empirical robustness without any theoretical guarantees, we minimize the maximum loss from the worst case adversarial examples: $\min_{\theta} \frac{1}{|D|} \sum_{x,y \in D} \max_{\|\delta\| \leq \epsilon_{rob}}\ell(\model(x+\delta), y)$~\cite{madry2018towards,trades}.
Certified robustness gives an upper bound on the loss for adversarial examples within a perturbation budget~\cite{certifiedRobustness2020sok,lecuyerCertified}. $\robustmodel$ is the adversarially trained, robust, variant of $\model$.

\item\label{rd2}\textbf{Outlier removal} safeguards against \ref{r2} by treating poisoning examples and backdoors as being closer to outliers of $\dtrain$'s distribution~\cite{borgnia2021strong}. 
\begin{arxiv}
For instance, in federated learning, model parameters corresponding to poisoned data samples are removed using robust aggregation at the server to obtain the final model~\cite{guerraoui2018hidden,poisondef}. 
\end{arxiv}
While these are empirical defences against \ref{r2}, certified robustness provides an upper bound on the prediction loss for poisoned data records in $\dtrain$~\cite{rab,crfl}. We refer to the model obtained after retraining after outlier removal as $\noOutlierModel$.

\item\label{rd3} \textbf{Watermarking} safeguards against \ref{r3}  by embedding out-of-distribution data (e.g., backdoors), referred to as ``watermarks'' in $\dtrain$~\cite{sablayrolles2020radioactive,watermarking,dawn,watermarkingadv}) %available only to model trainer 
which can be extracted later to demonstrate ownership. %Watermarks should transfer to $\derivedmodel$. 
$\watermarkModel$ is the watermarked variant of $\model$.

\item\label{rd4} \textbf{Fingerprinting} safeguards against \ref{r3} by comparing the fingerprints of $\model$ and $\derivedmodel$ to detect if $\derivedmodel$ was derived from $\model$. Unlike watermarking which requires retraining a model with watermarks, fingerprints are extracted from an already trained $\model$~\cite{lukas2021deep,caoFingerprint,peng2022fingerprinting,paramfingerprint,waheed2023grove}. Fingerprints can be generated from model characteristics (e.g., observables~\cite{paramfingerprint,waheed2023grove}), transferable adversarial examples~\cite{lukas2021deep,caoFingerprint,peng2022fingerprinting} or identifying overlapping $\dtrain$~\cite{maini2021dataset}.
\end{enumerate}

%To safeguard against privacy risks, we discuss: (a) differential privacy (\ref{pd1}) to mitigate \ref{p1} and \ref{p2}; (b) attribute and distribution privacy (\ref{pd2}) as potential mitigation against \ref{p3} and \ref{p4} respectively.
For privacy (\ref{pd1} and \ref{pd2}):
\begin{enumerate}[label=\textbf{PD\arabic*},leftmargin=*,leftmargin=*,wide,  labelindent=0pt]
\item\label{pd1} \textbf{Differential privacy}~\cite{dpsgd} (DP) says that given two neighboring training datasets $\dtrain$ and $\dtrain'$ differing by one record, a mechanism $\mathcal{M}$ preserves $(\epsilon_{dp}, \delta_{dp})$-DP for all $O \subseteq Range(\mathcal{M})$ if $Pr[\mathcal{M}(\dtrain) \in O] \le Pr[\mathcal{M}(\dtrain') \in O] \times e^\epsilon_{dp} + \delta$ where $\epsilon_{dp}$ is the privacy budget and $\delta_{dp}$ is probability mass of events where the privacy loss is larger than $e^\epsilon_{dp}$. For ML models, we add carefully computed noise to gradients during stochastic gradient descent updates, referred to as DPSGD~\cite{dpsgd}. This reduces the influence of any single data record in $\dtrain$ and thereby mitigates \ref{p1} and \ref{p2}~\cite{ye2022one}. We refer to differentially private variant of $\model$ obtained from DPSGD as $\privatemodel$.

\item\label{pd2} \textbf{Attribute privacy and PD3 distribution privacy} have been proposed as formal frameworks to preserve privacy in the context of database queries~\cite{chendistributionprivacy,zhang2022attribute}. 
\begin{arxiv}
These frameworks add carefully computed noise to query responses to make them indistinguishable either for different values of sensitive attributes (for attribute privacy) or datasets sampled from different distributions (for distribution privacy).
\end{arxiv}
Adapting this framework to ML is an open problem~\cite{propinf6,propinf4}. Hence, we do not include these defenses in our discussions.
\end{enumerate}

%We now present defenses to safeguard against \ref{u}, namely, training with group fairness (\ref{ud1}) and using model explanations (\ref{ud2} hereafter referred to as explanations) to mitigate and detect discriminatory behaviour respectively.
Finally, for fairness (\ref{ud1} and \ref{ud2}):
\begin{enumerate}[label=\textbf{FD\arabic*},leftmargin=*,leftmargin=*,wide,  labelindent=0pt]
\item\label{ud1}\textbf{Group fairness} ensures that 
\begin{arxiv}
different sensitive subgroups are treated equally, i.e.,     
\end{arxiv}
the behaviour of $\model$ is equitable across different subgroups. Formally, given the set of inputs $\mathcal{X}$, sensitive attributes $\mathcal{S}$ and true labels $\mathcal{Y}$, we can train $\model$ to satisfy different fairness metrics (e.g., demographic parity, equality of odds and equality of opportunity~\cite{hardt2016equality}) such that $\model(X)$ is statistically independent of $S$. Group fairness can be achieved by pre-processing $\dtrain$, in-processing by modifying training objective function or post-processing $\model(X)$~\cite{fairSurvey,fairSurvey2}. We refer to fair variant of $\model$ obtained from group fairness constraints as $\fairmodel$.

\item\label{ud2} \textbf{Explanations} increase transparency in $\model$'s computation by releasing additional information along with $\model(x)$. These help identify biases in $\model$ by checking if relevant attributes are being memorized~\cite{kim2018interpretability,gradcam}. None of the explanations require retraining. We present three types of explanations used in our paper:
\begin{itemize}[leftmargin=*]
\item Attribution-based: For an input $x$, explanations $\phi(x)$ are a vector $x = (x_1, \cdots x_n)$ indicating the influence of each of the $n$ attributes to $\model(x)$. 
\item Influence-based: For an input $x$, explanations $\phi(x)$ are data records in $\dtrain$ which resulted in $\model(x)$~\cite{koh2017understanding,pruthi2020estimating}.
\item Recourse-based (counterfactual): For an input $x$, these explanations output a data record $x_c$ which indicates the minimal change from $x$ to get the favourable outcome.
\end{itemize}
\end{enumerate}

\noindent We summarize the notations in Appendix~\ref{sec:notations} (Table~\ref{tab:notations}).

%% file: 3framework.tex
\section{Framework}\label{sec:framework}

We introduce a framework to examine how defences interact with various risks.
Hence, we conjecture that when a defence is effective in $\model$, \emph{overfitting} (Section~\ref{sec:overfitting}) and \emph{memorization} (Section~\ref{sec:memorization}) are the underlying causes affecting other risks (validated in Section~\ref{sec:interactions}). We further explore underlying factors that influence them.
% \begin{arxiv}
We also establish the relationship between overfitting and memorization (Section~\ref{sec:relation}).
% \end{arxiv}

\subsection{Overfitting}\label{sec:overfitting}

ML classifiers are optimized to improve the \textit{accuracy} on $\dtrain$ while ensuring that it generalizes to $\dtest$, i.e., high accuracy on both $\dtrain$ and $\dtest$.
However, failing to generalize to $\dtest$ despite high accuracy on $\dtrain$ is \textit{overfitting}.

\noindent\textbf{\underline{Measuring Overfitting.}}\label{sec:metricsOverfitting} We use \textit{generalization error} ($\mathcal{G}_{err}$) to measure overfitting which is the gap between the expected loss on $\dtrain$ and $\dtest$~\cite{generr}: $\mathcal{G}_{err} =  \expected{\loss(\model(x),y)}{(x,y) \sim \dtest} - \expected{\loss(\model(x),y)}{(x,y) \sim \dtrain}$.
Since expected loss is inversely proportional to model accuracy, henceforth, we rewrite $\mathcal{G}_{err}$ as the difference in accuracy between $\dtrain$ and $\dtest$.

\noindent\textbf{\underline{Factors Influencing Overfitting.}} We present two factors which influence overfitting: 
\begin{enumerate*}[label=\roman*),itemjoin={,\xspace}] 
\item size of $\dtrain$, i.e., |$\dtrain$| (\ref{data1}) 
\item $\model$'s capacity (\ref{model1})~\cite{traditionalBiasVariance,biasvarianceNN,biasvarianceNN2,biasvarianceNNOverview}.
\end{enumerate*}
They can be understood from the perspective of bias and variance: bias is an error from poor hyperparameter choices for $\model$ and variance is an error resulting from sensitivity to changes in $\dtrain$~\cite{traditionalBiasVariance}. High bias can prevent $\model$ from adequately learning relevant relationships between attributes and labels, while high variance leads to $\model$ fitting noise in $\dtrain$. 
\begin{arxiv}
There is a trade-off between bias and variance -- ideally, we want to find the right balance with the appropriate choice of \ref{data1} and \ref{model1}.   
\end{arxiv}

\begin{enumerate}[label=\texttt{D1},leftmargin=*,leftmargin=*,wide,  labelindent=0pt]
\item \label{data1}\textbf{Size of $\dtrain$ (|$\dtrain$|)}. Increasing |$\dtrain$| reduces overfitting: during inference, the likelihood that $\model$ has encountered a similar data record in $\dtrain$ is higher. Lower variance results in better generalization. 
% We include two other factors |$\dtrain$|-related in \ref{data1}: the number of poisons and the number of minority data records.
\end{enumerate}
\begin{enumerate}[label=\texttt{M1},leftmargin=*,leftmargin=*,wide,  labelindent=0pt]
\item \label{model1}\textbf{$\model$'s capacity.} Models with large capacity (large number of parameters) can perfectly fit $\dtrain$~\cite{arpitMemorization}. Smaller models have higher bias and cannot capture all patterns in $\dtrain$.
\end{enumerate}

\subsection{Memorization}\label{sec:memorization}

Memorization is measured with respect to a \emph{single data record} in $\dtrain$.
\begin{arxiv}
Multiple data records can be memorized simultaneously, e.g. a small set of outliers. 
\end{arxiv}
We categorize memorization into:

\begin{enumerate}[leftmargin=*,leftmargin=*,wide,  labelindent=0pt]
    \item \textbf{Data Record Memorization.} For classifiers, $\model$ tends to memorize entire data records and their corresponding labels, especially for outliers, mislabeled records, or those belonging to long-tail of $\dtrain$'s distribution~\cite{feldmanLongTail,zhang2017understanding,arpitMemorization}. This \textit{label-only memorization} is essential for $\model$ to achieve high accuracy on $\dtest$~\cite{brownMemorization}. Furthermore, memorized data records exhibit a higher influence on observables. 
    
    For generative models, which have low overfitting, memorization of entire data records results in their verbatim replication as seen in language and text-to-image diffusion models~\cite{extractlm,secretSharer,extractDiffusion,memorizationGenerative,memorizationGenerative}. This is influenced by the presence of duplicate data records in $\dtrain$ and generative model's large capacity~\cite{extractDiffusion,secretSharer,kandpal2022deduplicating}. Several metrics such as \textit{exposure}, \textit{k-Eidetic memorization}, and \textit{counterfactual memorization} have been used for generative models.

    \item \textbf{Attribute Memorization.} $\model$ can selectively memorize specific attributes from $\dtrain$, which is essential for accurate predictions~\cite{stableFeatures,mahajan2021connection}. There are two types of attributes which can be memorized by $\model$: \textit{stable attributes}, which remain consistent despite changes in data distributions or \textit{spurious attributes} which change $\model$'s prediction with change in distribution. $\model$'s generalization is better when it prioritizes learning stable attributes~\cite{stableFeatures}.
\end{enumerate}

% In all these cases, while memorization is necessary to improve accuracy, some useless data records in $\dtrain$ are memorized which do not necessarily improve accuracy on $\dtest$~\cite{feldmanLongTail,brownMemorization}. 

\noindent\underline{\textbf{Measuring Memorization.}} As we focus on classifiers, we use \textit{label-only memorization} of a data record $z_i=(x_i,y_i) \in S$, where $S \subseteq \dtrain$, as the difference in probability $\mathcal{P}(\model(x_i)=y_i)$ with and without $z_i$ in $\dtrain$~\cite{feldmanLongTail,zhang2021counterfactual,carlini2022quantifying}. It is given by $\mathtt{mem}(\mathcal{A},S,i) := \mathcal{P}_{\model\leftarrow \mathcal{A}(S)}[\model(x_i) = y_i] - \mathcal{P}_{\model\leftarrow \mathcal{A}(S\setminus z_i)}[\model(x_i) = y_i]$. If $\mathtt{mem}$ is close to one, it is likely that $\model$ has memorized $z_i$.% and $\mathtt{mem}=0$ indicates no memorization. %, i.e., leave-one-out stable where there is no change in $\model$'s behaviour with and without $z_i$.
% Furthermore, $\mathtt{mem}$ is related to measuring influence of a data record in $\dtrain$ w.r.t a data record in $\dtest$: $\mathtt{mem}$ is the same as computing the influence of $z_i$ on itself~\cite{feldmanMemorizationInfluence,zhang2021counterfactual}.
% It is given as $\mathtt{infl}(\mathcal{A},S,i,j) := P_{\model\leftarrow \mathcal{A}(S)}[\model(x_j') = y_j'] - P_{\model\leftarrow \mathcal{A}(S\setminus z_i)}[\model(x_j') = y_j']$. , i.e., $\mathtt{mem}(\mathcal{A},S,i) = \mathtt{infl}(\mathcal{A},S,i,i)$~.

\noindent\underline{\textbf{Factors Influencing Memorization.}} Different characteristics of 
\begin{enumerate*}[label=\alph*),itemjoin={,\xspace}] 
\item $\dtrain$
\item training algorithm
\item $\model$,
\end{enumerate*}
influence memorization of data records. For \textbf{$\dtrain$}, we consider 
\begin{enumerate*}[label=\roman*),itemjoin={,\xspace}] 
\item tail length of $\dtrain$'s distribution (\ref{data2})
\item number of input attributes (\ref{data3})
\item $\model$'s priority of learning stable attributes (\ref{data4}).
\end{enumerate*} 
\begin{arxiv}
Now, we elaborate on each of these factors.
\end{arxiv}

\begin{enumerate}[label=\texttt{D\arabic*},leftmargin=*,leftmargin=*,wide,  labelindent=0pt]
\setcounter{enumi}{1}
\item\label{data2} \textbf{Tail length of $\dtrain$'s distribution.} 
\begin{arxiv}
Consider the distribution of the number of $\dtrain$ data records across different classes. This distribution can be anywhere between not having any tail (e.g., uniform distribution) to long-tailed. Uniform distribution indicates that $\dtrain$ is balanced across different classes.
For a \emph{long-tailed} distribution, the majority of the data records belongs to the classes from the ``head'' of the distribution, while data from classes from the tail is scarce~\cite{li2023adversarial,wu2021adversarial}.
Data records from the tail classes are either rare, or outliers, and thus, $\model$ is more likely to memorise them~\cite{feldmanLongTail}.
\end{arxiv}
Generally, an increase in the tail length, indicated by having more classes with fewer data records than the head of the distribution. This corresponds to higher memorization of atypical data records from the tail classes~\cite{truthserum}.
\item\label{data3} \textbf{Number of input attributes}.
The number of input attributes inversely correlates with the extent of memorization of individual attributes by $\model$.
\item\label{data4} \textbf{Priority of learning stable attributes.} When $\model$ prioritizes learning stable attributes over spurious attributes, it generalizes better, making it invariant to changes in other attributes, dataset, or distribution~\cite{stableFeatures,mahajan2021connection,hartmann2022distribution}.
On the contrary, prioritizing spurious attributes increases $\model$'s memorization. Measuring stable attributes is challenging as there are no ground truths. Prior works have used \emph{mean rank metric} to measure the distance between any two inputs over their learnt representation by $\model$~\cite{mahajan2021domain,mahajan2021connection}. Its value is small if some attributes between two inputs are the same which corresponds to stable attributes. Hence, a low mean rank means $\model$ prioritized learning stable attributes.
\end{enumerate}

For characteristics of training algorithms, we identify 
% (a) \textit{regularization} (\ref{obj1}), 
\begin{enumerate*}[label=\roman*),itemjoin={,\xspace}] 
\item \textit{curvature smoothness} (\ref{obj1}) 
\item \textit{distinguishability of observables (between data records, subgroups, datasets, and models)} (\ref{obj2})
\item \textit{distance of $\dtrain$ data records to decision boundary} (\ref{obj3})
% \item \textit{training configuration} (\ref{model2}).
\end{enumerate*} as potential factors underlying memorization.

\begin{enumerate}[label=\texttt{O\arabic*},leftmargin=*,leftmargin=*,wide,  labelindent=0pt]
\setcounter{enumi}{0}
\item\label{obj1} \textbf{Curvature smoothness.}
The convergence of the loss on different data records towards a minima depends on the smoothness of the objective function.
Since different data records have varying levels of difficulty in terms of learning, a less smooth curvature can lead to local minima around difficult records~\cite{fioretto2022differential,carlini2022the,truthserum}.
Consequently, these data records demonstrate distinct memorization patterns, evident by different influence on the observables.
\item\label{obj2} \textbf{Distinguishability of observables.}
Due to the varying levels of difficulty in learning different records, and thus variable memorization, $\model$ exhibits distinguishability in observables.
We identify three subtypes based on distinguishability in observables between 
\begin{enumerate*}[label=\texttt{O2.\arabic*},leftmargin=*,  labelindent=0pt,itemjoin={,\xspace}]
\item\label{obj2-dataset} datasets %\begin{arxiv}(e.g., data records inside and outside $\dtrain$)\end{arxiv}
\item\label{obj2-subgroups} subgroups %identified by some sensitive attribute (e.g., males and females)
\item\label{obj2-models} models
\end{enumerate*}.
\item\label{obj3} \textbf{Distance of $\dtrain$ data records to decision boundary.} Due to the varying difficulty to learn different data records, they demonstrate variable memorization~\cite{tran2021differentially,tran2021fairness,labelOnlyMIA}. Memorized data records are closer to the decision boundary~\cite{fioretto2022differential}. 
% \item\label{model2} \textbf{Training configuration}, e.g., number of epochs, or hyperparameters for different defences (e.g., $\epsilon_{dp}$ for DPSGD and $\epsilon_{rob}$ for adversarial robustness) influences both memorization and overfitting. For example, training $\model$ for more epochs increases both memorization and overfitting~\cite{secretSharer}. 
\end{enumerate}

For model characteristics, we consider \textit{capacity}. This is the same as \ref{model1}.
Increasing $\model$'s capacity generally increases memorization by fitting a more complex decision boundary~\cite{lee2021deduplicating,MIARobustness,extractlm,extractDiffusion}. While larger model sizes may reduce overfitting when the model is too simple to effectively generalize to the dataset, over-parameterized models generalize well. Hence, $\mathcal{G}_{err}$ does not change and an increase in model size correlates with higher memorization.

% Further, reducing the number or precision of parameters, decreases $\model$'s capacity thereby limiting memorization (as evidenced by the lower membership inference attack success)~\cite{dudduGecko,MIARobustness}.

% \begin{enumerate}[label=\texttt{M\arabic*},leftmargin=*,leftmargin=*,wide,  labelindent=0pt]
% \setcounter{enumi}{1}
% \item\label{model2} \textbf{Model capacity.} 
% \begin{arxiv}
% The  amount of information $\model$ can store is represented by its capacity. 
% \end{arxiv}
% %and thereby perfectly fit $\dtrain$~\cite{collins2017capacity}. 
% Increasing $\model$'s capacity generally increases memorization fitting a more complex decision boundary~\cite{lee2021deduplicating,MIARobustness,extractlm,extractDiffusion}. Further, reducing the number or precision of parameters, decreases $\model$'s capacity thereby limiting memorization (as evidenced by the lower membership inference attack success)~\cite{dudduGecko,MIARobustness}. %\ref{model2} is the same as \ref{model1}. %and influences both overfitting and memorization. 
% We hereafter refer to \ref{model2} as \ref{model1}.
% \end{enumerate}

We summarize the different factors in Table~\ref{tab:sumFactors}. 
\begin{arxiv}
    We discuss the completeness of the set of underlying causes and factors, as well as extensions of our framework in Section~\ref{sec:discussion}.
\end{arxiv}
\input{tables/tab_factors}
\begin{arxiv}
\subsection{Overfitting and Memorization}\label{sec:relation}

We now illustrate relationship between overfitting and memorization. 

\noindent\textbf{Illustration Setup.} We generate a synthetic dataset for binary classification. We train a simple neural network with two hidden layers with ten neurons each. We measure overfitting using $\mathcal{G}_{err}$ and memorization using average $\mathtt{mem}$ across all data records in $\dtrain$. We depict the relationships between overfitting and memorization in Figure~\ref{fig:cases}.

\noindent\textbf{Illustration.} In Figure~\ref{fig:cases} (a), we assume that data records in $\dtrain$ and $\dtest$ are linearly separable and drawn from similar distributions. To illustrate overfitting without memorization (Figure~\ref{fig:cases} (b)), we introduce noise to data records in $\dtest$ so that some of them cross the decision boundary. Conversely, for memorization without overfitting (Figure~\ref{fig:cases} (c)), we add noise to $\dtrain$ to bring the classes closer to the decision boundary, while still maintaining linear separability in $\dtest$. Finally, to illustrate both overfitting and memorization (Figure~\ref{fig:cases} (d)), we introduce noise to both $\dtrain$ and $\dtest$, simulating a scenario where $\dtest$ does not precisely match the distribution of $\dtrain$ for overfitting, while data records in $\dtrain$ belonging to different classes are located close to one another which are then memorized. 

\noindent\textbf{Relationship Cases.} We describe four cases on whether overfitting and memorization occur: (a) no overfitting and no memorization (case \ref{case1}), (b) overfitting but no memorization (case \ref{case2}), (c) no overfitting but memorization (case \ref{case3}), and (d) both overfitting and memorization (case \ref{case4}).

\begin{figure}[!htb]
    \centering
    \subfigure[No overfitting, no memorization]{\includegraphics[width=0.23\textwidth]{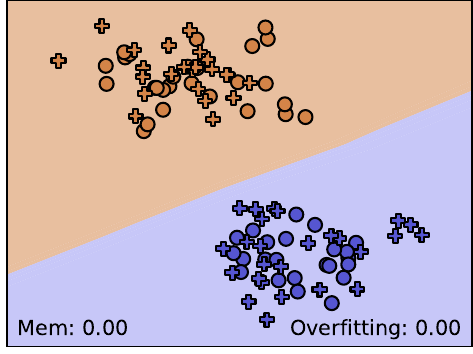}}
    \subfigure[Overfitting, no memorization]{\includegraphics[width=0.23\textwidth]{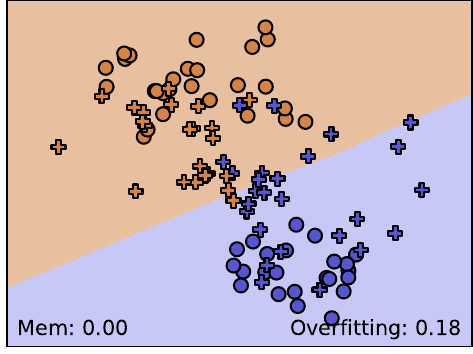}}\\
    \subfigure[Memorization, no overfitting]{\includegraphics[width=0.23\textwidth]{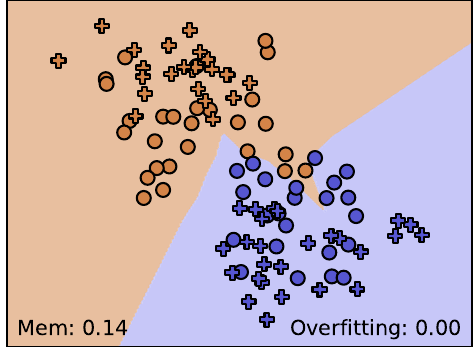}}
    \subfigure[Overfitting, memorization]{\includegraphics[width=0.23\textwidth]{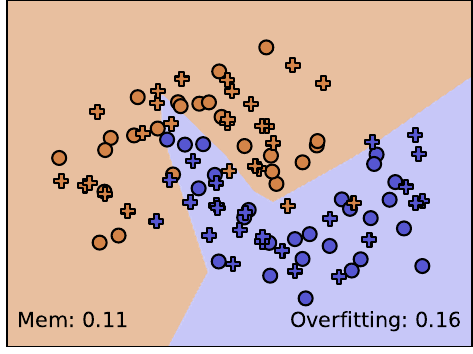}}
    \caption{Relationship between overfitting and memorization: Average $\mathtt{mem}$ across all data records in $\dtrain$ and overfitting ($\mathcal{G}_{err}$) are at the bottom. Circles indicate $\dtrain$ data records, and crosses indicate $\dtest$ data records.}
    \label{fig:cases}
\end{figure}

\begin{enumerate}[label=\texttt{C\arabic*},leftmargin=*,leftmargin=*,wide,  labelindent=0pt]

\item\noindent\textbf{No overfitting and no memorization (Figure~\ref{fig:cases}(a))}\label{case1} when data records in $\dtrain$ and $\dtest$ are linearly separable. Hence, $\model$ perfectly fits $\dtrain$ and generalizes to $\dtest$ ($\mathcal{G}_{err}=0$).
Further, data records in $\dtrain$ are far from the decision boundary, making them leave-one-out stable ($\mathtt{mem}$=0).

\item\noindent\textbf{Overfitting but no memorization (Figure~\ref{fig:cases}(b))}\label{case2} when $\dtrain$ is linearly separable but $\dtest$ is not. Here, $\model$ perfectly fits $\dtrain$ but does not generalize to $\dtest$. Additionally, data records in $\dtrain$ are far from the decision boundary making them leave-one-out stable ($\mathtt{mem}$=0).

\item\noindent\textbf{No overfitting but memorization (Figure~\ref{fig:cases}(c))}\label{case3} where data records in  $\dtrain$ and $\dtest$ are reasonably separable resulting in $\mathcal{G}_{err}=0$. However, with data records in $\dtrain$ being close to the decision boundary, we observe a high $\mathtt{mem}$.

\item\noindent\textbf{Overfitting and memorization (Figure~\ref{fig:cases}(d))}\label{case4} where $\model$ both overfits to $\dtrain$, i.e., $\model$ fits $\dtrain$ perfectly but does not generalize to $\dtest$, and memorizes some data records in $\dtrain$, i.e., they are not leave-one-out stable. In real-world applications, where $\dtrain$ and $\model$ are complex,  \ref{case4} is observed. Hence, hereafter we focus only on \ref{case4}. 
\end{enumerate}
\end{arxiv}

\begin{submit}
\subsection{Overfitting and Memorization}\label{sec:relation}

% The observation that memorization and overfitting are distinct and can occur simultaneously has been made previously in the context of generative models [42,22]. In Figure 1 we wanted to illustrate memorization and overfitting in the case of classifiers using our characterizations of them in Sections 3.1 and 3.2. Since whether or not they are independent is not critical to the rest of the paper, and in fact we are likely to encounter cases where both overfitting and memorization occur simultaneously (Figure 1 (d)), we will change our assertion to“Overfitting and Memorization are distinct and *can* occur independent[ly] of each other”
Memorization and overfitting have been shown as distinct phenomenon which can occur simultaneously as seen for generative models~\cite{secretSharer,memorizationGenerative}. We present four cases based on the presence or absence of overfitting and memorization in a model to illustrate the relation between them (Figure~\ref{fig:cases}). We use a synthetic dataset for binary classification and train a simple neural network with two hidden layers with ten neurons each. We measure overfitting using $\mathcal{G}_{err}$ and memorization using average $\mathtt{mem}$ across all $\dtrain$ data records.

\begin{figure}[!htb]
    \centering
    \subfigure[No overfitting, no memorization]{\includegraphics[width=0.23\textwidth]{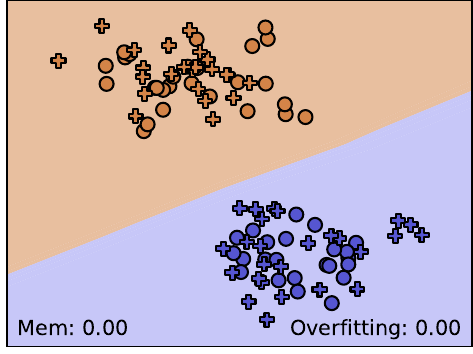}}
    \subfigure[Overfitting, no memorization]{\includegraphics[width=0.23\textwidth]{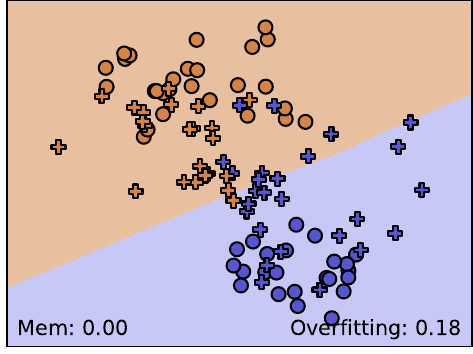}}\\
    \subfigure[Memorization, no overfitting]{\includegraphics[width=0.23\textwidth]{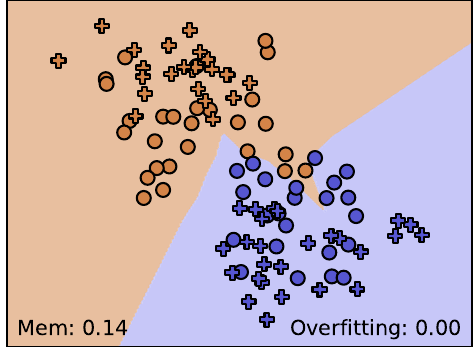}}
    \subfigure[Overfitting, memorization]{\includegraphics[width=0.23\textwidth]{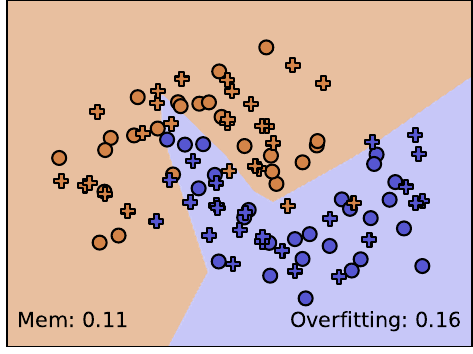}}
    \caption{Relationship between overfitting and memorization: Average $\mathtt{mem}$ across all data records in $\dtrain$ and overfitting ($\mathcal{G}_{err}$) are at the bottom. Circles indicate $\dtrain$ data records, and crosses indicate $\dtest$ data records.}
    \label{fig:cases}
\end{figure}

Figure~\ref{fig:cases}(a) is when data records in $\dtrain$ and $\dtest$ are linearly separable. Hence, $\model$ perfectly fits $\dtrain$ and generalizes to $\dtest$ ($\mathcal{G}_{err}=0$).
Further, data records in $\dtrain$ are far from the decision boundary ($\mathtt{mem}$=0). 
Figure~\ref{fig:cases}(b) is when $\dtrain$ is linearly separable but $\dtest$ is not ($\mathcal{G}_{err}>0$). However, data records in $\dtrain$ are far from the decision boundary ($\mathtt{mem}$=0). Figure~\ref{fig:cases}(c) is when data records in  $\dtrain$ and $\dtest$ are reasonably separable resulting in $\mathcal{G}_{err}=0$. However, with $\dtrain$ data records being close to the decision boundary, we observe a high $\mathtt{mem}$.
Figure~\ref{fig:cases}(d) is where $\model$ both overfits to $\dtrain$, i.e., $\model$ fits $\dtrain$ perfectly but does not generalize to $\dtest$, and memorizes some data records in $\dtrain$, i.e., they are not leave-one-out stable. This is generally seen in real-world applications, where $\dtrain$ and $\model$ are complex. Hence, hereafter we focus only on this setting. 
\end{submit}

%% file: tables/tab_factors.tex
\begin{table}[!htb]
\caption{Summary and description of different factors.}
\begin{center}
\footnotesize
\begin{tabular}{ c |  p{6.5cm} }
\hline
\textbf{Factor} & \textbf{Description}\\
% \hline
% \multicolumn{2}{c}{\textbf{Overfitting}}\\
\hline
\ref{data1} & Size of $\dtrain$ \\ %number of poisons or minority data records\\ 
% \hline
% \multicolumn{2}{c}{\textbf{Memorization}}\\
% \hline
\ref{data2} & Tail length of $\dtrain$'s distribution \\  
\ref{data3} & Number of input attributes \\
\ref{data4} & Priority of learning stable attributes\\
% \ref{obj1} & the presence of regularization for some protection mechanism\\
\ref{obj1} & Curvature smoothness of the objective function\\ %whether the interaction is consistent across linear and non-linear models\\
\ref{obj2} (\ref{obj2-dataset}) & Distinguishability of observables between datasets\\ %whether change in indistinguishability of the norms/predictions affects interaction\\
\ref{obj2} (\ref{obj2-subgroups}) & Distinguishability of observables across subgroups \\
\ref{obj2} (\ref{obj2-models}) & Distinguishability of observables across models\\
\ref{obj3} & Distance of $\dtrain$ data records to the decision boundary\\
% \hline
% \multicolumn{2}{c}{\textbf{Both}}\\
% \hline
% \ref{model2} & training configuration (e.g., number of epochs, $\epsilon_{dp}$, $\epsilon_{rob}$)\\
\ref{model1} & $\model$'s capacity\\
\hline
\end{tabular}
\end{center}
\label{tab:sumFactors}
\end{table}

%% file: 4.1protectionmech.tex
\section{Understanding Unintended Interactions}\label{sec:interactions}

We now show how factors influencing overfitting and memorization relate to defenses (Section~\ref{sec:revisitprotection}) and risks (Section~\ref{sec:revisitrisks}), summarizing in Table~\ref{tab:factorRelation}.
We then survey unintended interactions explored in prior work, and situate them within our framework (Section~\ref{sec:revisitinteractions}).
Finally, we suggest a guideline to help conjecture such interactions using our framework (Section~\ref{sec:guideline}).

\subsection{Revisiting Defenses for ML}\label{sec:revisitprotection}

We explore how the effectiveness of different defenses correlates with different factors.
We also indicate whether a defense degrades accuracy (\redsquare) or not (\greensquare).

\noindent\textbf{\ref{rd1} (Adversarial Training).}
$\model$ learns spurious attributes which are essential for achieving high accuracy and low $\mathcal{G}_{err}$.
However, this makes $\model$ susceptible to evasion~\cite{ilyas2019adversarial}.
$\robustmodel$ is compelled to learn stable attributes~\cite{tsipras2018robustness,ilyas2019adversarial}. %(\ref{data4}: $\uparrow$)
This improves robustness but with higher $\mathcal{G}_{err}$ and an accuracy drop (\redsquare)~\cite{ilyas2019adversarial}.
More training data can reduce overfitting~\cite{tsipras2018robustness}. %(\ref{data1}: $\uparrow$)
Increasing the tail length, lowers robustness for the tail classes~\cite{hu2022understanding,robustOddFair}. %(\ref{data2}: $\downarrow$)
Also, \ref{rd1} includes a regularization term to minimize the maximum loss from adversarial examples which results in a smoother curvature~\cite{madry2018towards}. %(\ref{obj1}: $\uparrow$)
This creates an explicit trade-off between generalization and robustness~\cite{szyller2022conflicting}. 
This is accompanied with an increase in the distance of $\dtrain$ data records to the decision boundary~\cite{Xu2023ExploringAE}. %(\ref{obj3}: $\uparrow$)
Further, $\robustmodel$ exhibits higher distinguishability between predictions on data records inside and outside $\dtrain$~\cite{MIARobustness}. %(\ref{obj2-dataset}: $\uparrow$)
Finally, $\robustmodel$ should have high capacity to learn adversarial examples~\cite{madry2018towards}. %(\ref{model1}: $\uparrow$)
%increases the gradient interpretability (\ref{obj2}: $\uparrow$)~\cite{robust2019private} -> for data reconstruction 

\noindent\textbf{\ref{rd2} (Outlier Removal).} Outliers are identified by measuring the influence of memorized data records on observables~\cite{jia2021scalability}. A longer tail increases the number of memorized outliers making their detection easy~\cite{wang2022partial}. %(\ref{data2}: $\uparrow$)
Some of these outliers contribute to $\model$'s accuracy~\cite{feldmanLongTail,feldmanMemorizationInfluence,brownMemorization}. Thus, removing such outliers increases overfitting in $\noOutlierModel$ which results in an accuracy drop (\redsquare)~\cite{jia2021scalability,jia19shapley,duddu2021shapr,jia2019knn}.

\noindent\textbf{\ref{rd3} (Watermarking).} Ideally, $\watermarkModel$ should maintain $\mathcal{G}_{err}$ similar to $\model$. In practice, $\watermarkModel$ incurs an accuracy drop (\redsquare)~\cite{watermarking,dawn,lukas2022sok}. A long tail enables stealthier watermarks from tail classes~\cite{watermarkbias}. %(\ref{data2}: $\uparrow$)
Lower distinguishability in observables for watermarks between $\model$ and $\derivedmodel$, but distinct from independently trained models, helps identifying $\derivedmodel$~\cite{watermarking}. %(\ref{obj2-models}: $\downarrow$)
High model capacity aids memorizing the watermarks~\cite{watermarking}. %(\ref{model1}: $\uparrow$)

\noindent\textbf{\ref{rd4} (Fingerprinting).} There is no explicit impact on overfitting, memorization, or accuracy (\greensquare) as \ref{rd4} does not require retraining. A lower distinguishability in fingerprints between $\model$ and $\derivedmodel$, but distinct from independently trained models, is preferred for effectively identifying $\derivedmodel$~\cite{waheed2023grove,lukas2021deep}. %(\ref{obj2-models}: $\downarrow$)
The effectiveness of fingerprints relies on $\model$'s overfitting and memorization. For instance, dataset inference~\cite{maini2021dataset} hinges on the distinguishability of observables between data records inside and outside $\dtrain$, for both $\model$ and $\derivedmodel$. This helps identify if they share overlapping $\dtrain$. Similarly, fingerprints based on adversarial examples rely on $\model$'s generalization for transferring adversarial examples to $\derivedmodel$~\cite{lukas2021deep,maini2021dataset,caoFingerprint,peng2022fingerprinting,wang2022role}. For effective transfer, a lower distance to the boundary is preferred~\cite{lukas2021deep,peng2022fingerprinting,caoFingerprint}. %(\ref{obj3}: $\downarrow$)

\noindent\textbf{\ref{pd1} (Differential Privacy).} Memorization of data records is reduced by lowering their influence on observables with \ref{pd1}. $\privatemodel$ is compelled to learn stable attributes as evident by better privacy guarantees in causal models~\cite{stableFeatures}. %(\ref{data4}: $\uparrow$)
Also, adding noise to gradients regularizes $\privatemodel$ which curbs overfitting but with an accuracy drop (\redsquare)~\cite{dpsgd,jayaraman2019evaluating,rahman2018membership}. 
This reduces distinguishability in observables for data records inside and outside $\dtrain$~\cite{dpsgd}, and decreases the distance of $\dtrain$ data records to the boundary~\cite{robustVDP1}. %(\ref{obj2-dataset}: $\downarrow$) %(\ref{obj3}: $\downarrow$)
Also, \ref{pd1} reduces the curvature smoothness~\cite{robustVDP1,robustVDP2}, and lower privacy for tail classes~\cite{DPLongTail,dpaccdisp}. %(\ref{data2}: $\downarrow$) %(\ref{obj1}: $\downarrow$)
Low capacity leads to reasonable privacy, without significant accuracy drop~\cite{shen2021towards}. %(\ref{model1}: $\downarrow$)

\noindent\textbf{\ref{ud1} (Group Fairness).} For fairness via in-processing algorithms, the regularization term explicitly creates a trade-off between fairness and accuracy~\cite{zhai2022understanding,veldanda2022fairness}, exacerbating overfitting with a drop in $\fairmodel$'s accuracy (\redsquare)~\cite{accfairtradeoff,rodolfa2021empirical,zhai2022understanding,veldanda2022fairness}. This reduces distinguishability in observables across subgroups~\cite{aalmoes2022leveraging,inprocessing1}. %(\ref{obj2-subgroups}: $\downarrow$)
Further, $\fairmodel$ prioritizes learning stable attributes which do not overlap with sensitive attributes~\cite{nabi2018fair,aalmoes2022leveraging}. %(\ref{data4}: $\uparrow$)
For equitable behaviour, $\fairmodel$ memorizes some data records in minority subgroups, thereby decreasing their distance to the boundary~\cite{miafairness,tran2022fairness}. %(\ref{obj3}: $\downarrow$)

\noindent\textbf{\ref{ud2} (Explanations).} There is no explicit impact on overfitting, memorization, or accuracy (\greensquare) as \ref{ud2} does not require retraining. More attributes increase the dimensions of the explanations, and reduce the influence of individual attributes which affects the quality of \ref{ud2}~\cite{miaexplanations}. %(\ref{data3}: $\downarrow$)
Also, \ref{ud2} exhibits distinguishability across subgroups~\cite{duddu2022inferring}. %(\ref{obj2-subgroups}: $\uparrow$)
The quality of explanations is tied to $\model$'s overfitting: they exhibit better approximation for data records in $\dtrain$ compared to $\dtest$~\cite{miaexplanations}. Therefore, \ref{ud2} exhibits higher distinguishability between data records inside and outside $\dtrain$~\cite{miaexplanations}. %(\ref{obj2-dataset}: $\uparrow$)

\begin{arxiv}
We summarize the relation between effectiveness of defenses with factors in Table~\ref{tab:factorRelation}. For a defense \protmech and a factor \factor, $\uparrow$ indicates \protmech positively correlates with \factor; $\downarrow$ indicates a negative correlation.
\end{arxiv}

%% file: 4.2risks.tex
\subsection{Revisiting Risks in ML}\label{sec:revisitrisks}

We now discuss different factors and how it correlates with the susceptibility on different risks.

\input{tables/tab_factorRelation}

\noindent\textbf{\ref{r1} (Evasion).} Longer tails increase \ref{r1}: adversarial examples from tail classes are more effective~\cite{wu2021adversarial,li2023adversarial}. %(\ref{data2}: $\uparrow$)
Lower curvature smoothness makes it easier to generate adversarial examples~\cite{madry2018towards}. %(\ref{obj1}: $\downarrow$)
Further, data records close to $\model$'s decision boundary require smaller perturbations to induce misclassification~\cite{karimi2019characterizing,chen2020hopskipjumpattack}. Hence, \ref{r1} risk decreases for data records farther away from the boundary~\cite{robustVDP1}. %(\ref{obj3}: $\downarrow$)
When overfitting and memorization are pronounced, the decision boundary is more complex and bigger, making \ref{r1} easier~\cite{chen2020hopskipjumpattack}.

\noindent\textbf{\ref{r2} (Poisoning).} 
Longer tail increases \ref{r2} as the poisons from tail classes are more discreet~\cite{paudice2018detection,poisondef,watermarkbias}. %(\ref{data2}: $\uparrow$)
Following watermarking literature which use poisons, the effectiveness of poisons is higher with larger $\model$ capacity due to their better memorization~\cite{watermarking}. % (\ref{model1}: $\uparrow$)

\noindent\textbf{\ref{r3} (Unauthorized Model Ownership).} Whitebox model theft does not explicitly change overfitting or memorization. Henceforth, we focus only on blackbox model extraction. \ref{r3} is less successful if $\model$ overfits or memorizes: accuracy of $\model$ is lower and consequently, labels from $\model$ for training $\derivedmodel$ are of a bad quality. Hence, $\derivedmodel$ has low accuracy~\cite{liu2022ml}.   
Also, for a given $\derivedmodel$, increasing $\model$'s capacity will decrease \ref{r3}~\cite{thievesSesame,orekondy2019knockoff}. %(\ref{model1}: $\downarrow$)

\noindent\textbf{\ref{p1} (Membership Inference).} Overfitting is correlated with an increase in \ref{p1}. Hence, \ref{p1} decreases with larger $|\dtrain|$ as it increases generalization~\cite{yeom,mia}. %(\ref{data1}: $\downarrow$)
Further, when $\model$ prioritizes learning stable attributes, \ref{p1} decreases due to better generalization~\cite{mahajan2021connection,stableFeatures}. %(\ref{data4}: $\downarrow$)
Also, a higher distinguishability between data records inside and outside $\dtrain$, increases \ref{p1}~\cite{mia,yeom,miaSurvey}. %(\ref{obj2-dataset}: $\uparrow$)
Memorized data records exhibit a higher influence on observables, rendering them more susceptible to \ref{p1}~\cite{duddu2021shapr,carlini2021membership}. 
Hence, longer tail increases memorized outliers thereby increasing their susceptibility~\cite{carlini2022the,carlini2021membership}. %(\ref{data2}: $\uparrow$)
Also, data records located farther from the decision boundary, with low memorization, are less susceptible to \ref{p1}~\cite{miaexplanations}. %(\ref{obj3}: $\downarrow$)
Finally, increasing $\model$'s capacity amplifies the memorization of $\dtrain$, which increases \ref{p1}~\cite{MIARobustness,dudduGecko}. %(\ref{model1}: $\uparrow$)

\noindent\textbf{\ref{p2} (Data Reconstruction).} Overfitting improves reconstruction of data records in $\dtrain$~\cite{nguyen2023re}. Higher distinguishability between records inside and outside $\dtrain$, increases \ref{p2}~\cite{genDataReconstruction}. %(\ref{obj2-dataset}: $\uparrow$)
Memorization amplifies the influence of data records on observables, leading to better reconstruction~\cite{remtoomuch,zhang2022model,wu2022linkteller,wen2022fishing}.
\change{\ref{p2}} is higher for outliers from tail classes~\cite{wen2022fishing}. %(\ref{data2}: $\uparrow$)
Further, more input attributes reduces accurate reconstruction of inputs~\cite{fredrikson1}. %(\ref{data3}: $\downarrow$)
Finally, higher distinguishability in observables across subgroups increases \ref{p2}~\cite{yang2020defending}. %(\ref{obj2-subgroups}: $\downarrow$)

\noindent\textbf{\ref{p3} (Attribute Inference).} Memorization of sensitive attributes is likely to increase \ref{p3}~\cite{truthserum}. Hence, longer tail increases memorized outliers, increasing their susceptibility~\cite{jayaraman2022attribute}. %(\ref{data2}: $\uparrow$)
Prioritizing learning stable attributes decreases \ref{p3}~\cite{mahajan2021connection,Song2020Overlearning}. %(\ref{data4}: $\downarrow$)
Higher distinguishability across subgroups increases \ref{p3}~\cite{aalmoes2022leveraging}. %(\ref{obj2-subgroups}: $\uparrow$)
Notably, overfitting has no discernible impact on this risk~\cite{liu2022ml}.

\noindent\textbf{\ref{p4} (Distribution Inference).} Empirical evidence suggests that overfitting reduces \ref{p4}, although the reason remains unexplained~\cite{propinf6,chaudhari2022snap}. 
Tail length correlates with \ref{p4}~\cite{chaudhari2022snap,chase2021property}. %(\ref{data2}: $\uparrow$)
Prioritizing stable attributes amplifies the risk because stable attributes are the same as the distributional property being inferred~\cite{propinf6}. %(\ref{data4}: $\uparrow$)
Higher distinguishability in observables across datasets with different distributional properties, increases \ref{p4}~\cite{propinf4,propinf6}. %(\ref{obj2-dataset}: $\uparrow$)
Finally, increasing $\model$'s capacity, without overfitting, increases \ref{p4}~\cite{chaudhari2022snap}. %(\ref{model1}: $\uparrow$)

\noindent\textbf{\ref{u} (Discriminatory Behaviour).} Overfitting makes bias prominent as evident by the accuracy disparity between minority and majority subgroups~\cite{dispvuln}. Increased memorization of minority subgroups contributes to their discrimination~\cite{miafairness}. Longer tails increase memorized outliers, worsening \ref{u}~\cite{fairSurvey}. %(\ref{data2}: $\uparrow$)
% because of $\model$ prioritizing spurious attributes~\cite{causalFairness}. %(\ref{data4}: $\downarrow$)
Increase in distinguishability of observables across subgroups correlates with increase in \ref{u}~\cite{inprocessing1,aalmoes2022leveraging}. %(\ref{obj2-subgroups}: $\uparrow$)

\begin{arxiv}
Table~\ref{tab:factorRelation} summarizes the above discussion. 
For a risk \risk and a factor \factor, $\uparrow$ indicates \risk positively correlates with \factor; $\downarrow$ indicates a negative correlation.
\end{arxiv}

%% file: tables/tab_factorRelation.tex
\begin{table*}[!htb]
\caption{Summary of correlation between effectiveness of a defense {\normalfont\protmech} and susceptibility to a risk {\normalfont\risk} with a factor {\normalfont\factor}:\\ $\uparrow$ indicates {\normalfont\protmech} or {\normalfont\risk} positively correlates with {\normalfont\factor}; $\downarrow$ indicates a negative correlation.
}
\begin{center}
\resizebox{\textwidth}{!}{%
\footnotesize
\begin{tabular}{ |p{8.5cm}|p{8.5cm}| } 
 \hline
\textbf{Defences (<$\uparrow$ or $\downarrow$>, \factor)} & \textbf{Risks (<$\uparrow$ or $\downarrow$>, \factor)} \\ 
\hline
\textbf{\ref{rd1} (Adversarial Training):}
\begin{itemize}[leftmargin=*]
    \item \ref{data1} $\uparrow$, $|\dtrain|$~\cite{tsipras2018robustness}
    \item \ref{data2} $\downarrow$, tail length~\cite{hu2022understanding,robustOddFair}
    \item \ref{data4} $\uparrow$, priority for learning stable attributes~\cite{tsipras2018robustness}
    \item \ref{obj1} $\uparrow$, curvature smoothness~\cite{madry2018towards}
    \item \ref{obj2-dataset} $\uparrow$, distinguishability in data records inside and outside $\dtrain$~\cite{MIARobustness}
    \item \ref{obj3} $\uparrow$, distance to boundary for most $\dtrain$ data records~\cite{Xu2023ExploringAE}
    \item \ref{model1} $\uparrow$, model capacity~\cite{madry2018towards}
\end{itemize} 

\textbf{\ref{rd2} (Outlier Removal):}
\begin{itemize}[leftmargin=*]
\item \ref{data2} $\uparrow$, tail length~\cite{wang2022partial} %memorization of new outliers from ``tail'' classes \textit{on retraining} % easier to detect outliers and remove them
\end{itemize}

\textbf{\ref{rd3} (Watermarking):}
\begin{itemize}[leftmargin=*]
\item\ref{data2} $\uparrow$, tail length~\cite{watermarkbias}
\item \ref{obj2-models} $\downarrow$, distinguishability in observables for watermarks between $\model$ and $\derivedmodel$, but distinct from independent models~\cite{watermarking}
\item\ref{model1} $\uparrow$, model capacity~\cite{watermarking}
\end{itemize}

\textbf{\ref{rd4} (Fingerprinting):}
\begin{itemize}[leftmargin=*]
\item \ref{obj2-models} $\downarrow$, distinguishability in observables for fingerprints between $\model$ and $\derivedmodel$, but distinct from independent models~\cite{waheed2023grove,lukas2021deep}
\item \ref{obj3} $\downarrow$, distance of $\dtrain$ data records to boundary~\cite{lukas2021deep,peng2022fingerprinting,caoFingerprint}
\end{itemize}

\textbf{\ref{pd1} (Differential privacy):}
\begin{itemize}[leftmargin=*]
\item \ref{data2} $\downarrow$, tail length~\cite{dpaccdisp,DPLongTail}
\item \ref{data4} $\uparrow$, priority for learning stable attributes~\cite{stableFeatures}
\item \ref{obj1} $\downarrow$, curvature smoothness~\cite{robustVDP1,robustVDP2}
\item \ref{obj2-dataset} $\downarrow$, distinguishability for data records inside and outside $\dtrain$~\cite{dpsgd}%gradient norm~\cite{robustVDP1} and 
% \item \ref{obj2-subgroups} $\downarrow$, distinguishability in observables across subgroups~\cite{ye2022one}
\item \ref{obj3} $\downarrow$, distance of $\dtrain$ data records to decision boundary~\cite{robustVDP1}
\item \ref{model1} $\downarrow$, model capacity~\cite{shen2021towards} %more noise needed lowers performance
\end{itemize}

\textbf{\ref{ud1} (Group Fairness):}
\begin{itemize}[leftmargin=*]
% \item \ref{data2} $\downarrow$, tail length~\cite{miafairness}
\item \ref{data4} $\uparrow$, priority for learning stable attributes~\cite{causalFairness,nabi2018fair,aalmoes2022leveraging}
\item \ref{obj2-subgroups} $\downarrow$, distinguishability in observables across subgroups~\cite{aalmoes2022leveraging,inprocessing1}
\item \ref{obj3} $\downarrow$, distance to decision boundary for most $
\dtrain$ data records~\cite{tran2022fairness}
\end{itemize}

\textbf{\ref{ud2} (Explanations):}
\begin{itemize}[leftmargin=*]
% \item\ref{data2} $\uparrow$ distinguishability across subgroups~\cite{explfairness1,explfairness2}
\item \ref{data3} $\downarrow$, number of input attributes~\cite{miaexplanations}
\item \ref{obj2-dataset} $\uparrow$, distinguishability in data records inside and outside $\dtrain$~\cite{miaexplanations}
\item \ref{obj2-subgroups} $\uparrow$, distinguishability in observables across subgroups~\cite{duddu2022inferring}
\end{itemize}

&

\textbf{\ref{r1} (Evasion):}
\begin{itemize}[leftmargin=*]
\item \ref{data2} $\uparrow$, tail length~\cite{wu2021adversarial,li2023adversarial}
% \item \ref{data4} $\downarrow$, priority for learning stable attributes~\cite{ilyas2019adversarial}
\item \ref{obj1} $\downarrow$, curvature smoothness~\cite{madry2018towards}
\item \ref{obj3} $\downarrow$, distance of $\dtrain$ data records to boundary~\cite{robustVDP1}
\end{itemize}

\textbf{\ref{r2} (Poisoning):}
\begin{itemize}[leftmargin=*]
\item \ref{data2} $\uparrow$, tail length~\cite{paudice2018detection,poisondef,watermarkbias}
\item \ref{model1} $\uparrow$, model capacity~\cite{watermarking}
% \item \ref{data4} $\uparrow$, priority for learning stable attributes~\cite{tao2021better}
% \item \ref{obj2} $\uparrow$ with larger gradient norms~\cite{hong2020effectiveness,poisonDP}
% poisons have a lrger gradient norm and hence can hide within the model's higher gradient norm for different inputs
\end{itemize}

\textbf{\ref{r3} (Unauthorized Model Ownership):}
\begin{itemize}[leftmargin=*]
% \item \ref{data2} $\uparrow$ with more query data~\cite{explanationsExtraction1,explanationsExtraction2}
% \item \ref{obj2} $\downarrow$ with noisy prediction confidence~\cite{dpextract2}
% \item \ref{obj2} $\uparrow$, distinguishability in observables between $\model$ and $\derivedmodel$~\cite{watermarking,waheed2023grove}
\item \ref{model1} $\downarrow$, model capacity~\cite{orekondy2019knockoff,thievesSesame}
% \item \ref{obj3} $\uparrow$ using queries closer to decision boundary~\cite{dpextract2}
\end{itemize}

\textbf{\ref{p1} (Membership Inference):}
\begin{itemize}[leftmargin=*]
\item \ref{data1} $\downarrow$, |$\dtrain$|~\cite{yeom,mia}
\item \ref{data2} $\uparrow$, tail length~\cite{carlini2022the,carlini2021membership}
\item \ref{data4} $\downarrow$, priority for learning stable attributes~\cite{mahajan2021connection,stableFeatures} 
\item \ref{obj2-dataset} $\uparrow$, distinguishability for data records inside and outside $\dtrain$~\cite{mia}
\item\ref{obj3} $\downarrow$, distance to decision boundary~\cite{miaexplanations}
\item \ref{model1} $\uparrow$, model capacity~\cite{MIARobustness,dudduGecko}
\end{itemize}

\textbf{\ref{p2} (Data Reconstruction):}
\begin{itemize}[leftmargin=*]
\item\ref{data2} $\uparrow$, tail length~\cite{wen2022fishing}
\item\ref{data3} $\downarrow$, number of input attributes~\cite{fredrikson1}
% \item\ref{data4} $\uparrow$, priority for learning stable attributes~\cite{robust2019private}
\item \ref{obj2-dataset} $\uparrow$, distinguishability for data records inside and outside $\dtrain$~\cite{genDataReconstruction,nguyen2023re}
\item\ref{obj2-subgroups} $\uparrow$, distinguishability in observables across subgroups~\cite{yang2020defending}
% \item \ref{obj2} $\uparrow$ with interpretable gradients~\cite{robust2019private}
\end{itemize}

\textbf{\ref{p3} (Attribute Inference):}
\begin{itemize}[leftmargin=*]
\item\ref{data2} $\uparrow$, tail length~\cite{jayaraman2022attribute}
\item\ref{data4} $\downarrow$, priority for learning stable attributes~\cite{mahajan2021connection,Song2020Overlearning}
\item\ref{obj2-subgroups} $\uparrow$, distinguishability in observables across subgroups~\cite{aalmoes2022leveraging}
\end{itemize}

\textbf{\ref{p4} (Distribution Inference):}
\begin{itemize}[leftmargin=*]
\item\ref{data2} $\uparrow$, tail length~\cite{chaudhari2022snap,chase2021property}
\item \ref{data4} $\uparrow$, priority for learning stable attributes~\cite{propinf6}
\item\ref{obj2-dataset} $\uparrow$, distinguishability in observables between datasets~\cite{propinf4,propinf6}
\item\ref{model1} $\uparrow$, model capacity~\cite{chaudhari2022snap}
\end{itemize}

\textbf{\ref{u} (Discriminatory behaviour):}
\begin{itemize}[leftmargin=*]
\item \ref{data2} $\uparrow$, tail length~\cite{fairSurvey}
% \item \ref{data4} $\downarrow$, priority for learning stable attributes~\cite{causalFairness}
\item \ref{obj2-subgroups} $\uparrow$, distinguishability in observables across subgroups~\cite{inprocessing1}
\end{itemize}
\\
 \hline
\end{tabular}
}
\end{center}
\label{tab:factorRelation}
\end{table*}

%% file: 4.3interactions.tex
\subsection{Surveying Unintended Interactions}\label{sec:revisitinteractions}

\input{tables/tab_interactions.tex}

We now survey various unintended interactions reported in the literature and situate them within our framework. 
We selected papers on various interactions using Google Scholar, focusing on those published in top-tier venues. Next, we examined their citations and related work to uncover additional works. We use existing surveys (e.g., Gittens et al.~\cite{surveyTradeoff}) to confirm that our coverage is reasonable.

We present the different interactions explored in prior works in Table~\ref{tab:interactions}.
We mark the type of interaction (indicated as ``\textbf{I}'') between \protmech and \risk as \redmark\xspace if ``an increase (decrease) in effectiveness of \protmech correlates with an increase (decrease) in \risk'' and \greenmark\xspace if ``an increase (decrease) in effectiveness of \protmech correlates with a decrease (increase) in \risk''. 
When \textbf{I} is \graymark, prior work has not studied this particular interaction and the middle columns are also left empty. When \textbf{I} is \redmark\xspace or \greenmark\xspace, empty middle columns correspond to cases where prior work merely identifies the type of interaction without evaluating the influence of the factors.

We use the different factors from Table~\ref{tab:sumFactors} to gain a more nuanced understanding of the interactions.
We mark a factor with \cmark if there are empirical results regarding its influence, \smark if there are theoretical results and \xmark if it is only conjectured. 
Exceptions to any interactions are listed under ``References'' and discussed in the text.

\noindent\textbf{\underline{\newtext{\ref{rd1} (Adversarial Training)}}}

\noindent\greenmark\xspace\textbf{\ref{r1} (Evasion)} is less effective with \ref{rd1} which is specifically designed to resist adversarial examples~\cite{madry2018towards,trades,certifiedRobustness2020sok}. Using the min-max objective function increases the curvature smoothness which reduces the possibility of generating adversarial examples~\cite{madry2018towards}. %(\ref{obj1}$\rightarrow$\cmark)
Further, \ref{rd1} pushes the decision boundary away from the data records~\cite{trades}. %(\ref{obj3}$\rightarrow$\cmark)
Also, $\robustmodel$ requires a high capacity to learn a complex decision boundary to fit adversarial examples~\cite{madry2018towards}. %(\ref{model1}$\rightarrow$\cmark)
Longer tail increases the effectiveness of adversarial examples from the tail classes despite using \ref{rd1}~\cite{li2023adversarial,wu2021adversarial}. %(\ref{data2}$\rightarrow$\cmark)
% This necessitates adjusting \ref{rd1} to ensure equitable robustness for all classes~\cite{li2023adversarial,wu2021adversarial}.

\noindent\redmark\xspace\textbf{\ref{r2} (Poisoning)} can be effective with \ref{rd1}~\cite{wen2023is}. Recall that $\robustmodel$ is forced to learn stable attributes.
Hence, it might seem that poisons, generated by manipulating spurious attributes, does not affect $\robustmodel$'s predictions~\cite{tao2021better}.
However, Wen et al.~\cite{wen2023is} design poisons to degrade $\robustmodel$'s accuracy by manipulating the stable attributes.

\noindent\redmark\xspace\textbf{\ref{r3} (Unauthorized model ownership)} can be more prevalent as model theft of $\robustmodel$ is easier than $\model$~\cite{khaled2022careful}. 
However, no reasons were explored. We conjecture that since $\robustmodel$ has uniform predictions on neighboring data records~\cite{juuti2019prada,stealml,madry2018towards,certifiedRobustness2020sok}, hence, $\derivedmodel$ requires fewer queries compared to $\model$. %(\ref{data1}$\rightarrow$\xmark) 
The above interaction assumes \adv is a malicious suspect. 
For a malicious accuser, \ref{rd1} can mitigate false claims~\cite{liu2023false}, which decreases \ref{r3} (\greenmark).

\noindent\redmark\xspace\textbf{\ref{p1} (Membership inference)} is easier with \ref{rd1}~\cite{MIARobustness,Hayes2020TradeoffsBM}. Recall that \ref{rd1} modifies $\robustmodel$’s decision boundary to memorize adversarial examples in $\dtrain$~\cite{MIARobustness}.
This increases the influence of some records on observables. Consequently, the distinguishability between data records inside and outside $\dtrain$ is higher~\cite{MIARobustness}. %\cmmnt{(\ref{obj2-dataset}$\rightarrow$\cmark)}
Hence, \ref{p1} increases.
Hayes et al.~\cite{Hayes2020TradeoffsBM} provably show that increasing $|\dtrain|$ lowers overfitting and hence, reduces \ref{p1}. %(\ref{data1}$\rightarrow$\smark,\cmark).
Further, \ref{p1} increases with $\robustmodel$'s capacity due to higher memorization. %\cmmnt{(\ref{model1}$\rightarrow$\cmark).}

\noindent\redmark\xspace\textbf{\ref{p2} (Data reconstruction)} is easier with \ref{rd1}~\cite{privacyFLrobust,robust2019private}:
$\robustmodel$ has interpretable gradients~\cite{robustVacc}, which enhances the quality of reconstruction.
% $\robustmodel$ with skip connections reduce the risk by averaging information available for \adv~\cite{robust2019private}.
Further, consistent attributes across inputs, are easier to reconstruct.%(\ref{data4}$\rightarrow$\xmark) 
Finally, lower $\robustmodel$'s capacity increases \ref{p2}~\cite{privacyFLrobust}. %(\ref{model1}$\rightarrow$\cmark)

\noindent\redmark\xspace\textbf{\ref{p4} (Distribution inference)} is easier as $\robustmodel$ tends to learn stable attributes which matches with distributional properties being inferred~\cite{propinf6}. %(\ref{data4}$\rightarrow$\xmark)
However, increasing $\epsilon_{rob}$ reduces \ref{p3} as $\robustmodel$'s accuracy decreases.

\noindent\redmark\xspace\textbf{\ref{u} (Discriminatory behaviour)} increases with \ref{rd1} as evident from the accuracy disparity of $\robustmodel$ across different classes~\cite{hu2022understanding,robustOddFair,croce2020robustbench}.
Longer tail contributes to the tail classes having lower robustness~\cite{hu2022understanding,robustOddFair}. %(\ref{data2}$\rightarrow$\smark,\cmark)
Also, increasing $\epsilon_{rob}$, increases the class-wise disparity~\cite{ma2022on}.
Empirically, $|\dtrain|$, $\robustmodel$'s capacity, type of $\dtrain$ (e.g., images, tabular), and $\robustmodel$'s architecture and optimization algorithms \emph{do not} influence the interaction~\cite{hu2022understanding,robustOddFair}.

\noindent\textbf{\underline{\newtext{\ref{rd2} (Outlier Removal)}}}

\noindent\greenmark\xspace\textbf{\ref{r1} (Evasion)} is less effective with \ref{rd2}~\cite{grosse2017statistical}. Adversarial examples are outliers compared to $\dtrain$'s distribution which can detected before passing them as inputs~\cite{grosse2017statistical}.

\noindent\greenmark\xspace\textbf{\ref{r2} (Poisoning)} is less effective with \ref{rd2}~\cite{poisonSurvey,paudice2018detection}. Poisons have a different distribution from $\dtrain$ and can be detected and removed as outliers during pre-processing~\cite{paudice2018detection}. During training, the influence of outliers, often correspond to poisons, on observables can be minimized~\cite{poisondef}.
% As expected, increasing the number of poisons can effectively reduce the accuracy thereby increasing $\mathcal{G}_{err}$~\cite{paudice2018detection}.

\noindent\redmark\xspace\textbf{\ref{p1} (Membership inference)} is arguably worse with \ref{rd2} followed by retraining~\cite{carlini2022the}.
Memorized outliers are more susceptible to membership inference. %(\ref{data2}$\rightarrow$\cmark)
However, after removing these outliers, $\noOutlierModel$ memorizes other data records making them more susceptible than before~\cite{carlini2022the}.
%Furthermore, it was observed that 
Varying $|\dtrain|$, $\noOutlierModel$'s capacity, or the number of duplicates in $\dtrain$ does \emph{not} have any impact on the risk~\cite{carlini2022the}.

\noindent\redmark\xspace\textbf{\ref{p3} (Attribute inference)} is more effective as retraining after removal of memorized outliers constituting the long-tail of $\dtrain$, makes new data records susceptible~\cite{jayaraman2022attribute}. %(\ref{data2}$\rightarrow$\cmark)

\noindent\redmark\xspace\textbf{\ref{u} (Discriminatory behaviour)} increases as outliers could constitute a statistical minority. %(\ref{data2}$\rightarrow$\xmark)
This exacerbates bias against minorities by excessively identifying them as outliers~\cite{outlierfair}. On increasing $|\dtrain|$, more minority data records are classified as outliers further exacerbating the bias~\cite{outlierfair}. %(\ref{data1}$\rightarrow$\cmark)

\noindent\textbf{\underline{\newtext{\ref{rd3} (Watermarking)}}}

\noindent\redmark\xspace\textbf{\ref{r2} (Poisoning)} is used by most watermarking schemes thereby requiring $\model$ to be susceptible to it~\cite{poisonOverfit,watermarking,watermarkInversion,li2022untargeted}. These schemes inject data records closer to outliers for memorization by $\model$. Hence, longer tails can help generate effective and discreet poisons as watermarks. %(\ref{data2}$\rightarrow$\xmark)

\noindent\greenmark\xspace\textbf{\ref{r3} (Unauthorized model ownership)} can be detected with \ref{rd3}~\cite{dawn,watermarking,lukas2022sok}. $\watermarkModel$ should have enough capacity to memorize watermarks. Therefore, increasing capacity can improve effectiveness of watermarking~\cite{lukas2022sok}. %(\ref{model1}$\rightarrow$\cmark)
Also, longer tails help generate more discreet watermarks. %(\ref{data2}$\rightarrow$\xmark) 

\noindent\redmark\xspace\textbf{\ref{p1} (Membership inference)}, \textbf{\ref{p2} (Data reconstruction)}, \textbf{\ref{p3} (Attribute inference)}, \textbf{\ref{p4} (Distribution inference)} do not have any prior evaluations with watermarks.
However, prior work has explored stealthy poisoning where mislabeled data records are added to $\dtrain$ to increase different privacy risks without an accuracy drop~\cite{truthserum,chen2022amplifying,chaudhari2022snap,chase2021property}.
We liken this form of poisoning to \ref{rd3}~\cite{poisonOverfit,watermarking,watermarkInversion,li2022untargeted}, and use their observations to justify different interactions.

We conjecture that adding watermarks from tail classes, similar to poisoning, will increase privacy risks. Hence, a longer tail is likely to increase different privacy risks. %(\ref{data2}$\rightarrow$\xmark) 
Further, memorization of watermarks increases their influence on observables. This increases the distinguishability in observables across datasets, subgroups, and models. %\ref{obj2-dataset}, \ref{obj2-subgroups}, \ref{obj2-models}$\rightarrow$\cmark) 
% This makes $\watermarkModel$ susceptible to different privacy risks. 
Moreover, the decision boundary is more complex, consequently increasing the susceptibility of outliers closer to the decision boundary~\cite{truthserum,chen2022amplifying}. %(\ref{obj3}$\rightarrow$\cmark) 
For distribution inference, the risk increases with lower capacity, and higher $|\dtrain|$~\cite{chaudhari2022snap,chase2021property}. %(\ref{model1}$\rightarrow$\cmark, \ref{data1}$\rightarrow$\smark,\cmark)

\noindent\redmark\xspace\textbf{\ref{u} (Discriminatory behaviour)} could worsen with watermarking schemes which rely on adding bias to $\dtrain$~\cite{watermarkbias}. Watermarks (with incorrect labels) belonging to the tail classes are more effectively memorized for ownership verification~\cite{watermarkbias,adversarialBias,jagielski2021subpopulation}. Hence, increasing the tail length is likely to increase the effectiveness of watermarks. %(\ref{data2}$\rightarrow$\xmark)

\noindent\textbf{\underline{\newtext{\ref{rd4} (Fingerprinting)}}}

\noindent\redmark\xspace\textbf{\ref{r1} (Evasion)} can be used for fingerprinting with adversarial examples, necessitating $\model$'s susceptibility to this risk~\cite{lukas2021deep,peng2022fingerprinting,caoFingerprint}. Hence, when $\model$ is robust against evasion, fingerprints are less effective~\cite{lukas2021deep}. Such fingerprints require $\dtrain$ records to be closer to class boundaries~\cite{caoFingerprint}. %(\ref{obj3}$\rightarrow$\cmark)
Further, overfitting and memorization make the decision boundary complex, which helps generate fingerprints. To transfer these fingerprints from $\model$ to $\derivedmodel$, the distinguishability between their observables should be low but distinct from independent models~\cite{lukas2021deep,peng2022fingerprinting,caoFingerprint}. %(\ref{obj2-models}$\rightarrow$\cmark)

\noindent\greenmark\xspace\textbf{\ref{r3} (Unauthorized model ownership)} can be detected using fingerprints. Fingerprints based on adversarial examples are closer to the outliers in the tail classes of $\dtrain$'s distribution. 
%Hence, longer tail and smaller distance of $\dtrain$ data records to the boundary helps to generate effective fingerprints~\cite{lukas2021deep,peng2022fingerprinting,caoFingerprint,maini2021dataset}. %(\ref{data2}$\rightarrow$\cmark)  %(\ref{obj3}$\rightarrow$\cmark) 
% Such fingerprints are easier to generate when s is small~\cite{maini2021dataset}.
Unique intermediate activations for a given input can also be used as a fingerprint which have low distinguishability between $\model$ and $\derivedmodel$ but distinct from independent models~\cite{waheed2023grove}. %(\ref{obj2-models}$\rightarrow$\cmark)

\noindent\redmark\xspace\textbf{\ref{p1} (Membership inference)} can be used to construct fingerprints which relies $\model$ being susceptible to it.
Dataset inference~\cite{maini2021dataset} uses a weak membership inference attack to identify $\model$'s decision boundary, that constitute its fingerprint. %(\ref{obj3}$\rightarrow$\cmark)
Specifically, it distinguishes between the distribution of distances of data records in $\dtrain$ and any other independent dataset to the decision boundary, and \emph{not} any individual data records.

\noindent\textbf{\underline{\newtext{\ref{pd1} (Differential Privacy)}}}

\noindent\redmark\xspace\textbf{\ref{r1} (Evasion)} is easier for $\privatemodel$.
DPSGD reduces the distance between $\dtrain$ records and decision boundaries making it easier to generate adversarial examples~\cite{robustVDP1}. %(\ref{obj3}$\rightarrow$\cmark)
Also, training with DPSGD results in a less smooth curvature which can can potentially increase \ref{r1}~\cite{robustVDP1,moosavi2019robustness,robustVDP2,advattTransfer}. %(\ref{obj1}$\rightarrow$\cmark) 
$\privatemodel$ creates a false sense of robustness via gradient masking, i.e., modifying the gradients to be useless for gradient based evasion (as observed in~\cite{robustVDP1}). However, $\privatemodel$ is still susceptible to non-gradient evasion. Bu and Zhang~\cite{bu2023differentially} show that training with DPSGD can result in adversarial robustness depending on the choice of hyperparameters (\greenmark). Further exploration is required to understand the conditions under which DPSGD and adversarial robustness align.
    
\noindent\redmark\xspace\textbf{\ref{r2} (Poisoning)} can be more effective against $\privatemodel$. Prior works have suggested that \ref{r2} decreases as \ref{pd1} reduces their influence on the observables~\cite{hong2020effectiveness,poisonDP}:
DPSGD constrains gradient magnitudes which enhances $\privatemodel$'s robustness against poisons.
However, it has been shown that the choice of hyperparameters (e.g., noise multiplier, gradient clipping) are the reason for an improvement in robustness~\cite{jagielskiDP}. More investigation is required to understand the root cause of this interaction.

\noindent\greenmark\xspace\textbf{\ref{p1} (Membership inference)} is less effective on $\privatemodel$ due to \ref{pd1}'s regularizing effect as evident from low $\mathcal{G}_{err}$~\cite{dpsgd}. This reduces the distinguishability between data records inside and outside $\dtrain$~\cite{dpsgd,jayaraman2019evaluating,liu2022ml}. %(\ref{obj2-dataset}$\rightarrow$\cmark) 
Further, \ref{pd1} provably bounds \ref{p1} for i.i.d. data~\cite{jayaraman2019evaluating,rahman2018membership}.
Prioritizing causal attributes gives a higher privacy guarantee while reducing the risk to \ref{p1}~\cite{stableFeatures}. %(\ref{data4}$\rightarrow$\smark,\cmark)
However, \ref{pd1}'s guarantee does not hold for \ref{p1} on non-i.i.d data~\cite{humphries2020differentially}. Also, duplicates in $\dtrain$ weaken the bound and the achievable $\epsilon_{dp}$ is too high for \ref{p1}.

\noindent\greenmark\xspace\textbf{\ref{p2} (Data reconstruction)} is less effective against $\privatemodel$ as \ref{pd1} reduces the distinguishability between data records inside and outside $\dtrain$~\cite{ye2022one}. %(\ref{obj2-dataset}$\rightarrow$\cmark) 
The lower quality of information available to \adv reduces \ref{p2}.

\noindent\greenmark\xspace\textbf{\ref{p3} (Attribute inference)} was less effective against $\privatemodel$~\cite{liu2022ml}, but no explanation was presented.
%However, no reason was explored as to why the risk decreases.
We conjecture that \ref{pd1}, by reducing the memorization of sensitive attributes and their influence on observables, reduces the distinguishability across subgroups~\cite{jia2021scalability}. %(\ref{obj2-subgroups}$\rightarrow$\xmark)

\noindent\greenmark\xspace\textbf{\ref{p4} (Distribution inference)} is less effective assuming \adv does not know $\privatemodel$'s training hyperparameters to adapt the attack~\cite{propinf6}.
\ref{pd1} has a regularizing effect which reduces $\privatemodel$'s ability to learn $\dtrain$'s distribution. This reduces the distinguishability of observables across datasets with different distributional properties thereby reducing \ref{p4}~\cite{propinf6}. %(\ref{obj2-dataset}$\rightarrow$\cmark)
However, if \adv is aware of training hyperparameters, they can adapt their attack to increase \ref{p4} (\redmark)~\cite{propinf6}.

\noindent\redmark\xspace\textbf{\ref{u} (Discriminatory behaviour)} increases for $\privatemodel$ as evident by accuracy disparity between minority and majority subgroups~\cite{dpaccdisp}.
This effect varies with $\privatemodel$'s architecture. %(\ref{model1}$\rightarrow$\cmark) 
Also, training hyperparameters (e.g., clipping and gradient noise) disproportionately affect underrepresented classes~\cite{tran2021differentially,esipova2023disparate}. Further, longer tail results in poor privacy protection for the tail classes~\cite{DPLongTail}. These data records have a high gradient norm and are closer to the decision boundary, which exacerbates the disparity~\cite{tran2021differentially}. %(\ref{data2}$\rightarrow$\xmark, \ref{obj3}$\rightarrow$\cmark)
Finally, lowering $\epsilon_{dp}$ makes the disparity worse~\cite{dpaccdisp}.

\noindent\textbf{\underline{\newtext{\ref{ud1} (Group Fairness)}}}

\noindent\redmark\xspace\textbf{\ref{r1} (Evasion)} is easier against $\fairmodel$. Data records in $\dtrain$ have a smaller distance to the decision boundary making it easier to generate adversarial examples~\cite{tran2022fairness}. %(\ref{obj3}$\rightarrow$\cmark) 
Robustness, on the other hand, is higher for data records with a larger distance to the decision boundary.
Hence, the optimization for \ref{ud1} is contrary to the robustness requirement~\cite{tran2022fairness}. %(\ref{obj1}$\rightarrow$\smark,\cmark)

\noindent\redmark\xspace\textbf{\ref{r2} (Poisoning)} is more effective against $\fairmodel$~\cite{poisonFair,adversarialBias,poisoningFairML,Mehrabi2021}.
It aims to increase $\fairmodel$'s disparity, and differs in ways to craft poisons which include: using $\fairmodel$'s gradients~\cite{poisonFair}, distorting $\dtrain$'s distribution~\cite{adversarialBias}, skewing the decision boundary~\cite{Mehrabi2021}, and maximizing the covariance between the sensitive attributes and labels~\cite{Mehrabi2021}. 

\noindent\redmark\xspace\textbf{\ref{p1} (Membership inference)} is more effective against $\fairmodel$~\cite{miafairness,tian2023fd}. Memorization of some data records lowers their distance to decision boundary thereby increasing their influence on observables which increases \ref{p1}~\cite{miafairness}. %(\ref{obj3}$\rightarrow$\cmark) 
Further, \ref{p1} increases with $\fairmodel$'s capacity and $|\dtrain|$ assuming the fraction of minority subgroup remains the same~\cite{miafairness}. %(\ref{data1}, \ref{model1}$\rightarrow$\cmark)
This is assuming equalized odds as a fairness metric. It is not clear how the nature of the interaction changes for a different fairness metric (e.g., demographic parity).

\noindent\greenmark\xspace\textbf{\ref{p3} (Attribute inference)} is less effective. $\fairmodel$ has lower distinguishability in predictions across subgroups~\cite{aalmoes2022leveraging}. %(\ref{obj2-subgroups}$\rightarrow$\smark, \cmark) 
Hence, \ref{ud1} specifically with demographic parity as a constraint, can provably and empirically reduce \ref{p3}~\cite{aalmoes2022leveraging}. On the contrary, Ferry et al.~\cite{ferry2023exploiting} show that it is possible to infer sensitive attributes from $\fairmodel$ which suggests an increase in \ref{p3} (\redmark). However, this is under a restrictive threat model where \adv knows the group fairness algorithm and hyperparameters to adapt their attack.

\noindent\redmark\xspace\textbf{\ref{p4} (Distribution inference)} is less effective with over-sampling and under-sampling data records. %(\ref{data1}$\rightarrow$\cmark) 
However, the disparity between different subgroups increases~\cite{propinf6}. Hence, the resistance to \ref{p4} and \ref{ud1} are in conflict.

\noindent\greenmark\xspace\textbf{\ref{u} (Discriminatory behaviour)} is less for $\fairmodel$. This can be achieved by \textit{pre-processing} $\dtrain$ to remove bias~\cite{preprocessing}. %(\ref{data1}$\rightarrow$\cmark)
Also, \textit{in-processing} using regularization reduces the distinguishability of predictions across subgroups to reduce disparity~\cite{inprocessing1,inprocessing12}. %(\ref{obj2-subgroups}$\rightarrow$\cmark)
Finally, \textit{post-processing} can help calibrate the predictions to satisfy some fairness metric~\cite{postprocessing}. 

\noindent\textbf{\underline{\newtext{\ref{ud2} (Explanations)}}}

\noindent\redmark\xspace\textbf{\ref{r1} (Evasion)} is easier as \ref{ud2} uses gradients which captures $\model$'s decision boundary. This helps generating adversarial examples, which in addition to forcing a misclassification, produces \adv's chosen explanation~\cite{underfire,quan2022amplification}.

\noindent\redmark\xspace\textbf{\ref{r2} (Poisoning)} is easier as \ref{ud2} provides a new attack surface -- $\model$ is forced to output \adv's target explanations to hide its discriminatory behaviour ~\cite{poisonInterpret,baniecki2021fooling,heo2019fooling,dombrowski2019explanations}.
Poisons shift the local optima, skew $\theta$ and $\model$'s decision boundary~\cite{poisonInterpret,heo2019fooling,dombrowski2019explanations,slack2021counterfactual}. Further, \ref{r2} is more effective with more capacity and smoother curvature~\cite{baniecki2021fooling,heo2019fooling,slack2021counterfactual}. %(\ref{model1}$\rightarrow$\cmark, \ref{obj1}$\rightarrow$\smark,\cmark)
Furthermore, explanations such as saliency maps and counterfactuals, can be similarly manipulated~\cite{heo2019fooling,slack2021counterfactual}. Less number of attributes reduces \ref{r2} by reducing the information available to \adv. %(\ref{data3}$\rightarrow$\cmark)

\noindent\redmark\xspace\textbf{\ref{r3} (Unauthorized model ownership)} is easier as \ref{ud2} provides information about $\model$'s decision boundary~\cite{explanationsExtraction1,explanationsExtraction2,quan2022amplification}.
For instance, exploiting counterfactuals results in $\derivedmodel$ with high accuracy and fidelity with fewer queries than exploiting predictions~\cite{explanationsExtraction2}.
The variability in \ref{r3} increases with \adv's dataset size and imbalanced dataset~\cite{explanationsExtraction2}.
The query efficiency can be improved by using a ``counterfactual of a counterfactual'' to train $\derivedmodel$~\cite{explanationsExtraction1}.

\noindent\redmark\xspace\textbf{\ref{p1} (Membership inference)} is easier as \ref{ud2} leaks membership status~\cite{miaexplanations,pawelczyk2022privacy,quan2022amplification,liu2024please}: explanations reflect the distance of data records to the boundary which makes them distinguishable for records inside and outside $\dtrain$. %(\ref{obj2-dataset}, \ref{obj3}$\rightarrow$\cmark)
Further, \ref{p1} decreases when there are fewer attributes but increases with |$\dtrain$|, and $\model$'s capacity~\cite{pawelczyk2022privacy}. %(\ref{data3}$\rightarrow$\cmark) %(\ref{data1}, \ref{model1}$\rightarrow$\cmark)

\noindent\redmark\xspace\textbf{\ref{p2} (Data reconstruction)} is easier with releasing \ref{ud2}~\cite{zhao2021exploiting}: explanations are a proxy for $\model$'s sensitivity to different attribute values which allow for accurate reconstruction -- the risk increases with the quality of explanations.
Moreover, some algorithms release influential $\dtrain$ data records for a given prediction on some input. This allows for perfect reconstruction of $\dtrain$ specially for data with a longer tail and with fewer attributes~\cite{miaexplanations}. %(\ref{data2}$\rightarrow$\smark,\cmark) (\ref{data3}$\rightarrow$\cmark)

\noindent\redmark\xspace\textbf{\ref{p3} (Attribute inference)} is more effective as \ref{ud2} leaks the values of sensitive attributes~\cite{duddu2022inferring}. The higher distinguishability across subgroups is a potential contributor. %(\ref{obj2-subgroups}$\rightarrow$\xmark)

\noindent\redmark\xspace\textbf{\ref{u} (Discriminatory behaviour)} can be detected using \ref{ud2} by identifying whether the relevant attributes are learnt by $\model$~\cite{gradcam,kim2018interpretability}. This suggests that \ref{ud2} reduces \ref{u} (\greenmark). However, explanations are flawed, i.e., they exhibit discriminatory behaviour across subgroups(\redmark)~\cite{explfairness1,explfairness2}. %(\ref{data2}$\rightarrow$\cmark)
This is more pronounced for large capacity~\cite{explfairness2}. %(\ref{model1}$\rightarrow$\cmark)
Also, explanation quality varies with the number of attributes~\cite{explfairness1}. %(\ref{data3}$\rightarrow$\cmark)

%% file: tables/tab_interactions.tex
% {\setlength\tabcolsep{4pt} 
\begin{table*}[!htb]
\caption{\underline{Unintended interactions between defenses with different risks}. \textbf{Type of interaction (I)}: \redmark $\rightarrow$ risk increases; \greenmark $\rightarrow$ risk decreases; \graymark $\rightarrow$ unexplored. We indicate exceptions to these under ``References'' and discuss them in the text. \textbf{Underlying factors:} \cmark $\rightarrow$ empirical; \smark $\rightarrow$ theoretical and \xmark $\rightarrow$ conjectured. Under \ref{obj2}, we write \ref{obj2-dataset}, \ref{obj2-subgroups}, and \ref{obj2-models} as \hyperref[obj2-dataset]{1}, \hyperref[obj2-subgroups]{2}, and \hyperref[obj2-models]{3} respectively.
We mark the factors evaluated for conjectured interactions as \conjmark (See Section~\ref{sec:conjectures}).}\label{tab:interactions}
\begin{center}
\resizebox{\textwidth}{!}{%
\footnotesize
\begin{tabular}{ p{3.3cm}|p{4.5cm}||c||P{0.7cm}|P{0.7cm}|P{0.7cm}|P{0.7cm}|P{0.7cm}|P{0.7cm}|P{0.7cm}|P{0.7cm}||p{3.3cm} }
\bottomrule

\toprule
\textbf{Defenses}& \textbf{Risks} & \textbf{I} & \multicolumn{1}{c|}{\textbf{OVFT}} & \multicolumn{5}{c|}{\textbf{Memorization}} & \multicolumn{2}{c||}{\textbf{Both}} & \textbf{References}\\
&  & & \ref{data1} &\ref{data2} & \ref{data3} & \ref{data4} & \ref{obj1} & \ref{obj2}  & \ref{obj3}  & \ref{model1} &  \\
\bottomrule

\toprule
\multirow{8}{*}{\textbf{\ref{rd1} (Adversarial Training)}} 
& \textbf{\ref{r1} (Evasion)} & \greenmark &  & \cmark &  &  & \cmark &  & \cmark  & \cmark & \cite{trades,madry2018towards,li2023adversarial,wu2021adversarial}\\
& \textbf{\ref{r2} (Poisoning)}  & \redmark &  &  &  & & & &  & & \cite{wen2023is,tao2021better} \\
& \textbf{\ref{r3} (Unauthorized Model Ownership)}  & \redmark & \xmark &  & & & & & & & \cite{khaled2022careful} (\cite{liu2022ml}: \greenmark)\\ % training data distribution of stolen model similar to adversarial training data distribution; attack success depends on size of training dataset
& \textbf{\ref{p1} (Membership Inference)}  & \redmark & \smark, \cmark &  &  & &  & \hyperref[obj2-dataset]{1}: \cmark &  &  \cmark & \cite{MIARobustness,Hayes2020TradeoffsBM}\\ %size of training data, noise added to training data records (feature distribution), model capacity, distance to decision boundary (higher sensitivity), regularization cause overfitting; 
& \textbf{\ref{p2} (Data Reconstruction)}  & \redmark &  &  & & \xmark  & & &  & \cmark & \cite{privacyFLrobust,robust2019private}\\ %batch size; model capacity and gradients to invert; feature types also play a role; model architecture (skip conections); variable activations (hence gradients)
& \textbf{\ref{p3} (Attribute Inference)}   & \graymark &  & & & & & & & & \\
& \textbf{\ref{p4} (Distribution Inference)}  & \redmark & &  & & \xmark & & &  & &  \cite{propinf6}\\
&  \textbf{\ref{u} (Discriminatory Behaviour)}  & \redmark & & \smark, \cmark &  & & & &   & & \cite{robustOddFair,croce2020robustbench,hu2022understanding,ma2022on}\\ %long tail; feature distribution; distance to decision boundary; learning process
\midrule

\multirow{8}{*}{\textbf{\ref{rd2} (Outlier Removal)}} 
& \textbf{\ref{r1} (Evasion)}  & \greenmark & & &  & & & & & &\cite{grosse2017statistical}\\
& \textbf{\ref{r2} (Poisoning)}  & \greenmark &  & &  & &  &  &  &  & \cite{poisonSurvey}\\
& \textbf{\ref{r3} (Unauthorized Model Ownership)}  & \graymark & & & & & & & & \\
& \textbf{\ref{p1} (Membership Inference)}  & \redmark & & \cmark & & &  &  & & & \cite{carlini2022the,duddu2021shapr}\\
& \textbf{\ref{p2} (Data Reconstruction)}  & \graymark & &  & & & & & & & \\ % outliers are more vulnerable; removing outliers make new records vulnerable to data reconstruction; similar to mia \cite{carlini2022the,duddu2021shapr} and attribute inference \cite{jayaraman2022attribute}
& \textbf{\ref{p3} (Attribute Inference)}   & \redmark &  & \cmark & & &  & & & & \cite{jayaraman2022attribute}\\  %different number of records for majority and minority class; 
& \textbf{\ref{p4} (Distribution Inference)}  & \graymark & & & & & & & & \\
&  \textbf{\ref{u} (Discriminatory Behaviour)}  & \redmark & \cmark & \xmark & & & & & & & \cite{outlierfair}\\ % size of data records in different subgroups;
% & \ref{u2} Incomprehension & & & & & & & & & & & & \\
\bottomrule

\multirow{8}{*}{\textbf{\ref{rd3} (Watermarking)}}
& \textbf{\ref{r1} (Evasion)} & \graymark & & & & & & & & & \\ %watermarks use adversarial examples during but the training maintains accuracy; impact of evasion attacks is not clear
& \textbf{\ref{r2} (Poisoning)}  & \redmark &  & \xmark & & & & & & & \cite{poisonOverfit,watermarking,watermarkInversion,li2022untargeted}\\
% &  &  & &  & &  &  & &  & \cite{poisonOverfit,dawn}\\ %decision boundary is modified by training with poisons; differential gradients/predictions which make it possible to detect watermarks; these poisons constitute long tail of model; regularization lowers success of watermark; higher model capacity allows for successful watermark; number of poisons impacts watermark success
& \textbf{\ref{r3} (Unauthorized Model Ownership)}    & \greenmark &  & \xmark & &  &  & \hyperref[obj2-models]{3}: \cmark & \cmark & & \cite{dawn,watermarking,lukas2022sok}\\
& \textbf{\ref{p1} (Membership Inference)}  & \redmark &  & \xmark & & & & \hyperref[obj2-dataset]{1}: \cmark & \cmark & & \cite{truthserum,chen2022amplifying}\\ %inferring whether some watermarked data records were used to train a model; watermarks modify the long tail of the distribution and feature distribution; ensure modification to decision boundary and variability in gradients to effectively infer data records; 
& \textbf{\ref{p2} (Data Reconstruction)} & \redmark &  & \xmark & & & & \hyperref[obj2-dataset]{1}: \cmark& \cmark & & \cite{truthserum}\\ 
& \textbf{\ref{p3} (Attribute Inference)}   & \redmark &  & \xmark & & & & \hyperref[obj2-subgroups]{2}: \cmark & \cmark & & \cite{truthserum}\\
& \textbf{\ref{p4} (Distribution Inference)}   & \redmark & \smark, \cmark & \xmark & & & & \hyperref[obj2-dataset]{1}: \cmark & \cmark & \cmark &  \cite{chaudhari2022snap,chase2021property}\\
&  \textbf{\ref{u} (Discriminatory Behaviour)}   & \redmark & & \xmark & & & & & & & \cite{watermarkbias}\\ % easily detected as outliers; out of distribution; triggers added as features; counts as minorty subgroup
\bottomrule

\multirow{8}{*}{\textbf{\ref{rd4} (Fingerprinting)}}
& \textbf{\ref{r1} (Evasion)}  & \redmark & & & & & & \hyperref[obj2-models]{3}: \cmark & \cmark &  & \cite{lukas2021deep,caoFingerprint,peng2022fingerprinting}\\ % Conferrable adversarial examples occur in those; adversarial subspaces where the decision boundary of surrogate and reference models differ; 
& \textbf{\ref{r2} (Poisoning)}   & \graymark & & & & & &  & & & \\ %balanced number of records in each class; decision boundary shift on adding watermarks; 
& \textbf{\ref{r3} (Unauthorized Model Ownership)}  & \greenmark &  & \cmark &  &  & & \hyperref[obj2-models]{3}: \cmark & \cmark &   & \cite{paramfingerprint,waheed2023grove,maini2021dataset,lukas2021deep}\\
& \textbf{\ref{p1} (Membership Inference)}   & \redmark & & & & & & & \cmark & & \cite{maini2021dataset}\\ %difference in predictions; using distnace from decision boundary; model architecture impact performance; number of data records for DI influences success
& \textbf{\ref{p2} (Data Reconstruction)}  & \graymark & & & & & & & & & \\
& \textbf{\ref{p3} (Attribute Inference)}   & \graymark  & & & & & & & & & \\
& \textbf{\ref{p4} (Distribution Inference)}  & \graymark & & & & & & & &  & \\
&  \textbf{\ref{u} (Discriminatory Behaviour)}  & \graymark & & & & & & & & & \\
\bottomrule

\multirow{8}{*}{\textbf{\ref{pd1} (Differential Privacy)}} 
& \textbf{\ref{r1} (Evasion)}  & \redmark  & & & & & \cmark & & \cmark &  & \cite{robustVDP1,robustVDP2} (\cite{bu2023differentially}:\greenmark)\\ % loss surface, gradients and decision boundary might have an effect
& \textbf{\ref{r2} (Poisoning)}  & \redmark  &  & & & & & & & & \cite{jagielskiDP,poisonDP,hong2020effectiveness}\\%poisoning modifies features which impacts gradients for clean and poison samples; % loss function plays a role wherein poison and clean samples have different loss trajectories; depends on the fraction of poisons inserted
& \textbf{\ref{r3} (Unauthorized Model Ownership)}  & \graymark & & & & &  & &  & &\\ %obfuscate decision boundary; adding noise to predictions to manipulate confidence scores making it harder to steal model
& \textbf{\ref{p1} (Membership Inference)}  & \greenmark &  &  &  & \smark, \cmark & & \hyperref[obj2-dataset]{1}: \cmark &  &  & \cite{jayaraman2019evaluating,rahman2018membership,liu2022ml,humphries2020differentially} \\
& \textbf{\ref{p2} (Data Reconstruction)}  & \greenmark &  &  &  &  & & \hyperref[obj2-dataset]{1}: \cmark &  &  & \cite{ye2022one,stock2023defending}\\
& \textbf{\ref{p3} (Attribute Inference)}  & \greenmark & & & & & & \hyperref[obj2-subgroups]{2}:  \xmark & & & \cite{liu2022ml}\\
& \textbf{\ref{p4} (Distribution Inference)}  & \greenmark & &  & & & & \cmark & & & \cite{propinf2,propinf6} (\cite{propinf6}: \redmark)\\
&  \textbf{\ref{u} (Discriminatory Behaviour)}   & \redmark & & \xmark & & & & & \cmark  & & \cite{dpaccdisp,esipova2023disparate,fioretto2022differential,tran2021differentially,zhang2020privacy}\\ %size of minority subgroups (long tailed), distance to decision boundary, non-linear objective function, variable gradients for different records
\bottomrule

\multirow{8}{*}{\textbf{\ref{ud1} (Group Fairness)}}
& \textbf{\ref{r1} (Evasion)}  & \redmark & &  & & & \cmark  & & \smark, \cmark & & \cite{tran2022fairness}\\
& \textbf{\ref{r2} (Poisoning)}  & \redmark & & &  & & &  &  &  & \cite{poisonFair,adversarialBias,poisoningFairML,Mehrabi2021}\\
& \textbf{\ref{r3} (Unauthorized Model Ownership)}  & \graymark  &  &  & & & & & & & \\
& \textbf{\ref{p1} (Membership Inference)}  & \redmark & \cmark & & &  &  &  & \cmark & \cmark & \cite{miafairness,tian2023fd}\\ %training dataset size; long tailed; model capacity; distance to decision boundary
& \textbf{\ref{p2} (Data Reconstruction)}  & \graymark &  &  & \conjmark & & & \hyperref[obj2-subgroups]{2}: \conjmark &  &  & Our conjecture: \greenmark\\ %size of training data, model architecture, variable gradents (as proxy for variable pmodel confidences)
& \textbf{\ref{p3} (Attribute Inference)}  & \greenmark & &  &  & & & \hyperref[obj2-subgroups]{2}:\smark,\cmark & & & \cite{aalmoes2022leveraging} (\cite{ferry2023exploiting}: \redmark) \\
& \textbf{\ref{p4} (Distribution Inference)}  & \redmark & \cmark & & & & & & & & \cite{propinf6}\\ 
&  \textbf{\ref{u} (Discriminatory Behaviour)}   & \greenmark &\cmark  & & & & & \hyperref[obj2-subgroups]{2}:\cmark &  &  &  \cite{preprocessing,inprocessing1,inprocessing12,postprocessing}\\
\midrule

\multirow{8}{*}{\textbf{\ref{ud2} (Explanations)}}
& \textbf{\ref{r1} (Evasion)}  & \redmark & &  & &  & & &  & & \cite{underfire,quan2022amplification} \\
& \textbf{\ref{r2} (Poisoning)}  & \redmark & &  & \cmark  & & \smark, \cmark &  & & \cmark & \cite{poisonInterpret,baniecki2021fooling,heo2019fooling,slack2021counterfactual}\\
& \textbf{\ref{r3} (Unauthorized Model Ownership)}  & \redmark & &  & & &  & &  &  & \cite{explanationsExtraction1,explanationsExtraction2,quan2022amplification}\\
& \textbf{\ref{p1} (Membership Inference)}  & \redmark & \cmark &  & \cmark & & & \hyperref[obj2-dataset]{1}: \cmark & \cmark & \cmark & \cite{miaexplanations,pawelczyk2022privacy,quan2022amplification,liu2024please}\\
& \textbf{\ref{p2} (Data Reconstruction)}  & \redmark &  & \smark, \cmark & \cmark &  & &  &  & & \cite{miaexplanations,zhao2021exploiting}\\
& \textbf{\ref{p3} (Attribute Inference)}  & \redmark & &  & & &  & \hyperref[obj2-subgroups]{2}: \xmark &  & & \cite{duddu2022inferring}\\ % variable distance to decision boundary, gradients (what explanations measure); features whose influence is computed is variabl3e for different records
& \textbf{\ref{p4} (Distribution Inference)}  & \graymark & & & \conjmark & &  & \hyperref[obj2-dataset]{1}: \conjmark & & \conjmark & Our Conjecture: \redmark\\ % explanations vary with different training data distribution; reveals information about decision boundary resulting in distinguishable gradients (hence explanations); makes it possible to infer properties
&  \textbf{\ref{u} (Discriminatory Behaviour)}  & \redmark & & \cmark &  \cmark & & &  &  & \cmark & \cite{explfairness1,explfairness2} (\cite{kim2018interpretability,gradcam}: \greenmark)\\
\bottomrule

\toprule
\end{tabular}
}
\end{center}
\end{table*}
% }

%% file: 4.4guideline.tex
\subsection{Guideline: Exploring Unintended Interactions}\label{sec:guideline}

We now present a guideline to identify the nature of an unintended interaction and the underlying factors influencing it. In Table~\ref{tab:factorRelation}, we summarized statements about how a change in: 
\begin{enumerate*}[leftmargin=*]
\item effectiveness of \protmech correlates with a change in \factor,
\item \factor correlates with \risk. 
\end{enumerate*}
%We indicate the two statements as ($\uparrow$ or $\downarrow$, $\uparrow$ or $\downarrow$).
We used $\uparrow$ to identify positive correlation and $\downarrow$ for negative correlation.
Hence, for a given \protmech and \risk with a common factor \commonfactor, we can infer how a change in effectiveness of \protmech correlates with \risk using the direction of the pair of arrows that describe how each of \protmech and \risk correspond to \commonfactor. 

Creating an exact algorithm for predicting unexplored interactions is difficult due to the intricate interplay of various factors. Therefore, we outline a guideline to make such conclusions with some expert knowledge about the nature of unintended interactions and underlying factors:
\begin{enumerate}[label=\texttt{G\arabic*},leftmargin=*,wide,labelindent=0pt]
\item\label{step1} For a given \protmech and \risk, identify \commonfactor{s} from Table~\ref{tab:factorRelation}.
\item\label{step2} For each \commonfactor, if change in \factor on using \protmech aligns with change in \risk, i.e., ($\uparrow$,$\uparrow$) or ($\downarrow$,$\downarrow$), then effectiveness of \protmech corresponds to an increase in \risk (\redmark) else \protmech negatively correlates with \risk (\greenmark) for ($\uparrow$,$\downarrow$) or ($\downarrow$,$\uparrow$).
\item\label{step3} Consider the suggestions from all \commonfactor{s} to conjecture the nature of the interaction. The final conclusion is:
\begin{enumerate*}[leftmargin=*]
\item common suggestion when all the \commonfactor agree on it,
\item the suggestion from the dominant \commonfactor determined using expert knowledge for conflicting suggestions from \commonfactor{s}.
\end{enumerate*}
\item\label{step4} Empirically evaluate to see if other non-dominant \factor{s} influence the interaction.
% \item\label{step5} \change{If there are no \commonfactor{s}, use empirical evaluation.}
\end{enumerate}

\noindent\textbf{Choosing Dominant \factor with Expert Knowledge.} Some factors, namely, \ref{obj1}, \ref{obj2}, \ref{obj3}, are exploited by different attacks. For instance, most inference attacks rely on \ref{obj2} while \ref{r1} (evasion) and \ref{r2} (poisoning) rely on \ref{obj1}, and \ref{obj3}. 
Using a defense affects these factors, which has a significant effect on the attack success. Therefore, we identify them as dominant compared to other factors which remain constant while evaluating if the risk changes on using a defense.

\noindent\textbf{Guideline's Coverage of Interactions.} To see how well our guideline covers different interactions, we compare conclusions from our guideline with corresponding observations from our survey as the ground truth (Section~\ref{sec:interactions}). The guideline and the ground truth are consistent in most cases.
We discuss a few exceptions in Section~\ref{sec:discussion}.
Next, we use our guideline to conjecture about unexplored interactions.

%% file: 5conjectures.tex
\section{Unexplored Interactions and Conjectures}\label{sec:conjectures}

We enumerate different unexplored interactions (\graymark\xspace in Table~\ref{tab:interactions}) in Table~\ref{tab:unexplored}. 
As all defenses and risks we surveyed hinge on factors related to overfitting and memorization (Section~\ref{sec:revisitprotection} and~\ref{sec:revisitrisks}), we conjecture that overfitting and memorization will influence these unexplored interactions.
To demonstrate the applicability of our framework, we apply our guideline from Section~\ref{sec:guideline} to present conjectures for two of these unexplored interactions and validate them empirically~\footnote{Code: \url{https://github.com/ssg-research/sok-unintended-interactions}.}
\begin{enumerate*}[label=\roman*),itemjoin={,\xspace}] 
\item Section~\ref{sec:fairreconstruct}: \ref{ud2} (explanations) and \ref{p4} (distribution inference)
\item Section~\ref{sec:propinfl}: \ref{ud1} (group fairness) and \ref{p2} (data reconstruction).
\end{enumerate*}
Among the unexplored interactions in Table~\ref{tab:interactions}, we selected two examples for empirical evaluation based on ease of implementation -- we already had convenient implementations for both attacks and defenses involved in these two examples.
We mark the factors evaluated for these interactions with \conjmark in Table~\ref{tab:interactions}.
These examples are intended for illustrating the applicability of our framework. We leave a systematic empirical exploration of unintended interactions as future work.
In Section~\ref{sec:remaining}, we speculate on the remaining interactions which we could not evaluate empirically.

\input{tables/tab_unexplored}

\subsection{\ref{ud1} \ding{222} \ref{p2}}\label{sec:fairreconstruct}

%We now investigate whether training $\model$ with group fairness corresponds to an increase or decrease in the data reconstruction.

\noindent\textbf{\underline{Conjecture.}} From Table~\ref{tab:factorRelation}, we identify \ref{obj2-subgroups} as a common factor (\commonfactor). As the arrows are in opposite directions, we conjecture that \ref{ud1} reduces \ref{p2} $\rightarrow$\greenmark. Now, we check for other factors: \ref{data3} suggests that \ref{p2} decreases anyway when there are more input attributes. Thus, we expect that the decrease due to \ref{ud1} is only evident when the number of attributes is small. Consequently, the difference in attack success between $\fairmodel$ and $\model$ becomes negligible. We describe the experimental setup below.

% We consider the reconstruction of $\fairmodel$'s inputs $x$, given $\fairmodel(x)$.
% Group fairness makes $\fairmodel(x)$ indistinguishable for different values of sensitive attributes~\cite{aalmoes2022leveraging}.
% It reduces the sensitivity of $\fairmodel$ to individual attributes, and thus the amount of information \adv can learn from $\fairmodel(x)$.
% Therefore, we conjecture that group fairness corresponds to a decrease in reconstruction attacks ($\rightarrow$\greenmark). 

\noindent\textbf{\underline{Threat Model.}}
We assume $x$ is sensitive and confidential.
\adv only observes $\fairmodel(x)$ for some unknown confidential $x$ which are to be reconstructed.
\adv also has access to auxiliary dataset which is split into $\dtrainadv$ to train an attack decoder model ($\attmodel$) and $\dtestadv$ for evaluation. 

\noindent\textbf{\underline{Attack Methodology.}}
\adv uses $\dtrainadv$ to train $\attmodel$ to map $\fairmodel(x)$ to $x$, minimizing the error between  $\attmodel(\fairmodelshadow(x))$ and the ground truth $x$.
Here, $\fairmodelshadow$ is a ``shadow model'' which is trained on \adv's auxiliary dataset to have similar functionality as $\fairmodel$.
We give the advantage to \adv and assume $\fairmodel == \fairmodelshadow$.
The attack is evaluated on $\dtestadv$.

\noindent\textbf{\underline{Setup.}}
We use the CENSUS dataset which consists of attributes like age, race, education level, to predict whether an individual’s annual income exceeds 50K. We select the top $10$ attributes and consider both race and sex attributes. We use $|\dtrain|$=15470, $|\dtrainadv|$=7735, and $|\dtestadv|$=7735. 

We train $\fairmodel$ with adversarial debiasing which uses an adversarial network $f_{inf}$ to infer sensitive attribute values from $\fairmodel(x)$~\cite{inprocessing1}.
The loss from $f_{inf}$ is used to train $\fairmodel$ such that $\fairmodel(x)$ is indistinguishable.

We measure fairness by (a) checking if p\%-rule is $>80\%$ as required by the regulations~\cite{pmlr-v54-zafar17a}, and (b) AUC score of $f_{inf}$ to correctly infer the sensitive attribute values (fair model should have a score of $0.50$).
To measure attack success, we use reconstruction loss ($l_2$-norm) between $x$ and $\attmodel(\fairmodelshadow(x))$ (higher the loss, less effective the attack).
We use a multi-layered perceptron (MLP) for $\fairmodel$ with hidden layer dimensions [32, 32, 32, 1], $f_{inf}$ with [32, 32, 32, 2], and $\attmodel$ with [18, 32, 64, 128, 10].

We compare the attack success with the baseline of data reconstruction against $\model$ with the same architecture as $\fairmodel$ but without group fairness. %The attack methodology described previously remains the same except for replacing $\model$ instead of $\fairmodel$. 

\noindent\textbf{\underline{Empirical Validation.}}
We validate our conjecture, show the effect on distinguishability across subgroups (\ref{obj2-subgroups}), and evaluate the influence of the number of input attributes (\ref{data3}).
% We leave a detailed evaluation of other factors as future work.

\noindent\textbf{Nature of Interaction.} We compare the reconstruction loss between $\fairmodel$ and $\model$ in Table~\ref{tab:fairInvType}. We confirm that training with adversarial debiasing is indeed fair (p\%-rule $>80$\% and AUC score $\sim$$0.50$). We observe that $\fairmodel$ has a higher reconstruction loss validating our conjecture.

\begin{table}[htb]
\caption{\textbf{Nature of Interaction.} Comparison of $\dtest$ accuracy, fairness (AUC scores, p\%-rule values) and reconstruction loss for $\model$ and $\fairmodel$ across ten runs. \\$\uparrow$: higher value is desirable and $\downarrow$: lower value is desirable.}
\begin{center}
\footnotesize
\begin{tabular}{ c | c | c   }
\hline
\textbf{Metric} & \textbf{$\model$} & \textbf{$\fairmodel$} \\
\hline
\textbf{Acc} $\uparrow$ & 84.40 $\pm$ 0.09 & 77.96 $\pm$ 0.58\\
\textbf{AUC} $\downarrow$ & 0.66 $\pm$ 0.00 & 0.52 $\pm$ 0.01\\
p\%-rule \textbf{\race} $\uparrow$ & 46.51 $\pm$ 0.59 & 93.69 $\pm$ 5.60\\
p\%-rule \textbf{\sex} $\uparrow$ & 29.70 $\pm$ 1.15  &  90.50 $\pm$ 6.87\\
\textbf{ReconLoss} $\uparrow$ & 0.85 $\pm$ 0.01 & 0.95 $\pm$ 0.02\\
\hline
\end{tabular}
\end{center}
\label{tab:fairInvType}
\end{table}

\begin{figure}[!htb]
    \centering
    \begin{minipage}[b]{\linewidth}
    \centering
    \subfigure[$\model$'s Prediction Distribution]{
    \includegraphics[width=\linewidth]{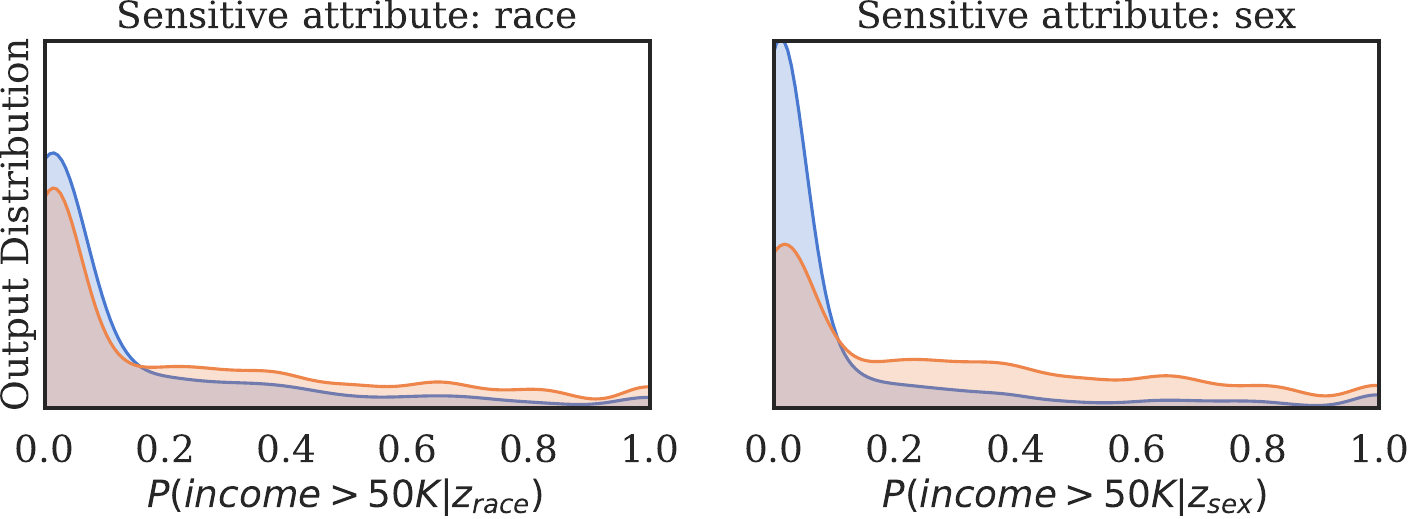}
    }
    \end{minipage}
    \begin{minipage}[b]{\linewidth}
    \centering
    \subfigure[$\fairmodel$'s Prediction Distribution]{
    \includegraphics[width=1\linewidth]{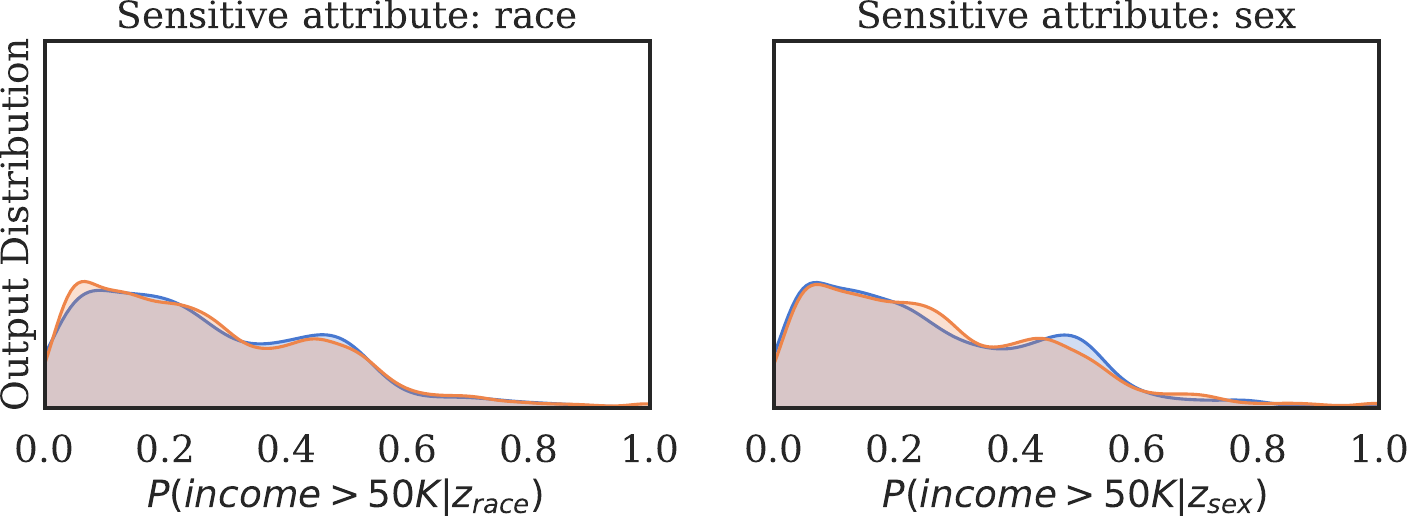}
    }
    \end{minipage}
    \caption{Distinguishability across subgroups (\ref{obj2-subgroups}) decreases for $\fairmodel$.}
    \label{fig:distribution}
\end{figure}

\noindent\textbf{Distinguishability in observables across subgroups (\ref{obj2-subgroups}).}
We plot $\fairmodel$'s prediction distribution for race (whites as {\color{red!40}red} vs non-whites as {\color{blue!40}blue}) and sex (males as {\color{red!40}red} vs females as {\color{blue!40}blue}) (Figure~\ref{fig:distribution}). 

We notice that there is no income disparity between whites-non-whites and males-females for $\fairmodel$.  %(\ref{obj2-subgroups}: \conjmark)
Further, $\fairmodel$'s predictions are normalized which reduces the information available to \adv for successful reconstruction.

\noindent\textbf{Number of Input Attributes (\ref{data3}).}
We present the impact of the number of input attributes in Table~\ref{tab:FairInvDimension}. 
Firstly, we note that the accuracy is similar across different number of input attributes for $\fairmodel$ regardless of the difference in reconstruction loss. Hence, the accuracy does not impact the risk, and only the number of input attributes is the variable.

{\setlength\tabcolsep{3pt}
\begin{table}[!htb]
\caption{\textbf{Number of Input Attributes (\ref{data3}).} Reconstruction loss and $\dtest$ accuracy for different number of input attributes for $\model$ and $\fairmodel$ (averaged across ten runs).}
\begin{center}
\footnotesize
\begin{tabular}{ c | c | c | c | c }
\hline
\textbf{\# Input} & \multicolumn{2}{c|}{\textbf{$\model$}} & \multicolumn{2}{c}{\textbf{$\fairmodel$}}\\ 
\textbf{Attributes} & \textbf{ReconLoss} & \textbf{Acc} & \textbf{ReconLoss} & \textbf{Acc}\\
\hline
\textbf{10} & 0.85 $\pm$ 0.00 & 84.40 $\pm$ 0.09 & 0.95 $\pm$ 0.02 & 78.96 $\pm$ 0.58\\
\textbf{20} & 0.93 $\pm$ 0.03 & 84.72 $\pm$ 0.22 & 0.93 $\pm$ 0.00 & 80.32 $\pm$ 1.12\\
\textbf{30} & 0.95 $\pm$ 0.02 & 84.41 $\pm$ 0.39 & 0.94 $\pm$ 0.00 & 79.50 $\pm$ 0.91\\
 \hline
\end{tabular}
\end{center}
\label{tab:FairInvDimension}
\end{table}
}

We present the reconstruction loss for a different number of input attributes \{$10, 20, 30$\}.
As expected, for $10$ attributes, group fairness reduces attack success significantly (as also seen in Table~\ref{tab:fairInvType}).
However, increasing the number of input attributes has no significant difference in reconstruction loss between $\model$ and $\fairmodel$, validating our conjecture. %(\ref{data3}: \conjmark).

\subsection{\ref{ud2} \ding{222} \ref{p4}}\label{sec:propinfl}

\noindent\textbf{\underline{Conjecture.}}
We identify \ref{obj2-dataset} as \commonfactor. As the arrows are the same direction, we conjecture that releasing \ref{ud2} increases \ref{p4} (\redmark).
Now, we check for other factors: the number of input attributes (\ref{data3}) and model capacity (\ref{model1}). For \ref{data3}, increasing the number of input attributes should reduce \ref{p4} by making it harder to capture distributional properties. For \ref{model1}, increasing $\model$'s capacity should increase memorization of distributional properties, thereby increasing \ref{p4}. We describe the experimental setup below.

\noindent\textbf{\underline{Threat Model.}}
We assume that on input $x$, $\model$ outputs both explanations $\phi(x)$ and $\model(x)$.
\adv differentiates between $\model$ trained on $\dtrain$ with ratio $\alpha_1$ vs. $\alpha_2$ (e.g., $50\%$ vs. $10\%$ women in $\dtrain$) by distinguishing between their respective explanations $\phi(x)^{\alpha_1}$, and $\phi(x)^{\alpha_2}$ using an ML attack model $\attmodel$. \adv has an auxiliary dataset which is split into $\dtrainadv$ and $\dtestadv$ to train and evaluate $\attmodel$. 

\noindent\textbf{\underline{Attack Methodology.}}
Given two ratios, $\alpha_1$ or $\alpha_2$, \adv samples $\dtrainadv$ to get $^{\alpha_1}\dtrainadv$ and $^{\alpha_2}\dtrainadv$ satisfying $\alpha_1$ or $\alpha_2$.
\adv then trains multiple \textit{shadow models} on these datasets which try to mimic $\model$; \adv uses $\phi'(x)^{\alpha_1}$ and $\phi'(x)^{\alpha_2}$ from the shadow models to train $\attmodel$.
Given $\model$ which is trained on $\dtrain$ satisfying some unknown ratio, \adv uses $\phi(x)$ where $x\in\dtestadv$ from $\model$ as input to $\attmodel$ to infer if $\dtrain$'s property was $\alpha_1$ or $\alpha_2$. 

\noindent\textbf{\underline{Setup.}} We use the CENSUS dataset from Section~\ref{sec:fairreconstruct} with $42$ input attributes.
We consider different ratios of women in $\dtrain$ as our property of interest in the range $0.1-0.9$ incremented by $0.1$.
Following prior work~\cite{propinf6,propinf4}, we set $\alpha_1=0.5$ while varying $\alpha_2$. For each value of $\alpha_2$, \adv trains $\attmodel$ to differentiate between $\alpha_1=0.5$ and the specific value of $\alpha_2$.
We measure the attack success using (distribution) inference accuracy, where $50\%$ is random guess.
We use an MLP with the following hidden layer dimensions: [1024, 512, 256, 128]. 
We use three explanation algorithms: IntegratedGradients~\cite{intgrad}, DeepLift~\cite{deeplift,deeplift2}, and SmoothGrad~\cite{smoothgrad}.
For explanations, we want to evaluate if releasing explanations increases \ref{p4}. Following the approach by Pawelczyk et al.~\cite{pawelczyk2022privacy}, assuming random guess (50\%) as the baseline, we want to show that an attack that \textit{only} makes use of explanations, can do better than the baseline. Our goal is to see if explanations leak distribution information, and not whether explanations can improve the state-of-the-art distribution inference attacks. 

\noindent\textbf{\underline{Empirical Validation.}}
We validate our conjecture and evaluate how it is influenced by the number of input attributes (\ref{data3}), and model capacity (\ref{model1}).
% We leave a detailed evaluation of other factors as future work.

\begin{figure}[!htb]
\includegraphics[width=0.6\linewidth]{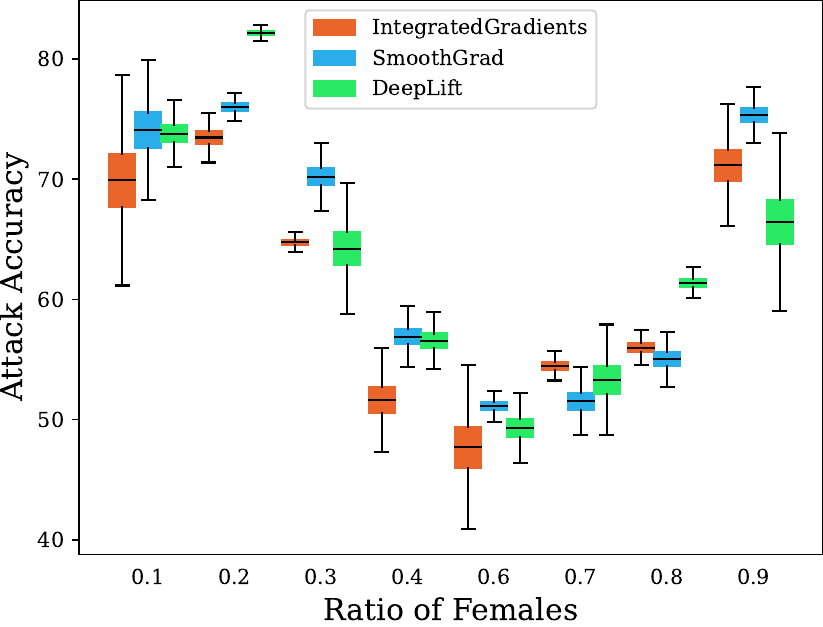}
\centering
\caption{\noindent\textbf{Nature of Interaction.} Accuracy to differentiate explanations from model trained on $\dtrain$ with $\alpha_1$=$0.5$ and $\alpha_2$ $\in$ \{$0.1-0.9$\}.}
\label{fig:explDistInf}
\end{figure}

\noindent\textbf{Nature of Interaction.}
We present the attack accuracy across different ratios and explanation algorithms in Figure~\ref{fig:explDistInf}. 
We use all 42 attributes for evaluation.
For the three explanation algorithms, the attack accuracy is higher than a random guess of 50\% for most of the ratios, validating our conjecture -- explanations indeed leak distribution properties of $\dtrain$, and increase \ref{p4}. 

\noindent\textbf{Number of input attributes (\ref{data3}).}
We use an attribute selection algorithm to select the top $35$, $25$ and $15$ attributes from the original $42$ attributes. We evaluate the impact of varying the top-k relevant attributes on the interaction. We fix $\alpha_1$=$0.5$ and $\alpha_2$=$0.1$ as it shows the highest attack accuracy (Figure~\ref{fig:explDistInf}).
We present the results in Table~\ref{tab:propinfdata3}.

{\setlength\tabcolsep{3pt} 
\begin{table}[htb]
\caption{\textbf{Number of input attributes (\ref{data3}).} Attack accuracy for $\alpha_1=0.5$ and $\alpha_2=0.1$ while varying the number of input attributes (results averaged across ten runs).}
\begin{center}
\footnotesize
\begin{tabular}{ c | c | c | c }
\hline
\textbf{\# Attributes}& \textbf{IntegratedGradients} & \textbf{DeepLift} & \textbf{SmoothGrad} \\ 
 \hline
\textbf{15} & 81.07 $\pm$ 2.13 & 78.74 $\pm$ 1.66 & 65.40 $\pm$ 1.39\\
\textbf{25} & 66.09 $\pm$ 0.95 & 73.64 $\pm$ 1.38 & 59.42 $\pm$ 1.09\\
\textbf{35} & 50.43 $\pm$ 0.59 & 59.93 $\pm$ 2.81 & 56.78 $\pm$ 1.93 \\
 \hline
\end{tabular}
\end{center}
\label{tab:propinfdata3}
\end{table}
}

We confirm that increasing the number of most relevant input attributes, decreases the attack success: %(\ref{data3}$\rightarrow$\conjmark) 
more number of attributes decreases the memorization of each attribute thereby reducing the quality of explanations.

\noindent\textbf{Model Capacity (\ref{model1}).}
We consider four MLPs with different model capacities: ``Model 1'' with dimensions [128], ``Model 2'' with [256, 128], ``Model 3'' with [256,512,256], and ``Model 4'' is the same as before. We present the attack accuracy in Table~\ref{tab:propinfModelCap}.

{\setlength\tabcolsep{3pt} 
\begin{table}[htb]
\caption{\textbf{Model capacity (\ref{model1}).} Attack accuracy on explanations corresponding to $\alpha_1$=0.5 and $\alpha_2$=0.1 for different models (averaged across ten runs).}
\begin{center}
\footnotesize
\begin{tabular}{ l | c | c | c }
\hline
\textbf{\# Parameters} & \textbf{IntegratedGradients} & \textbf{DeepLift} & \textbf{SmoothGrad}\\
 \hline
\textbf{Model 1 (5762)} & 47.57 $\pm$ 4.25 & 49.19 $\pm$ 2.75 & 53.26 $\pm$ 0.10\\
\textbf{Model 2 (44162)} & 53.29 $\pm$ 3.65 & 50.86 $\pm$ 3.24 & 62.40 $\pm$ 0.95\\
\textbf{Model 3 (274434} & 62.60 $\pm$ 2.74 & 67.73 $\pm$ 1.69 & 70.21 $\pm$ 0.73\\
\textbf{Model 4 (733314)} & 69.90 $\pm$ 3.24 & 73.78 $\pm$ 1.03 &  74.09 $\pm$ 2.17\\
 \hline
\end{tabular}
\end{center}
\label{tab:propinfModelCap}
\end{table}
}

All models have an accuracy of $82.10 \pm 1.17\%$ and close to perfect $\dtrain$ accuracy.
Hence, neither accuracy nor $\mathcal{G}_{err}$ impact \ref{p4}.
We observe that \ref{p4} increases with capacity: %(\ref{model1}: \conjmark)
a higher capacity correlates with greater memorization of $\dtrain$'s distributional properties.

\subsection{Other Unintended Interactions}\label{sec:remaining}

\begin{arxiv}
We now reflect on other unintended interactions.
\end{arxiv}
\noindent\textbf{\ref{rd1} (Adversarial Training) and \ref{p3} (Attribute Inference).} 
Here, \ref{data2} and \ref{data4} are the common factors. As the arrows are in the opposite directions, both suggest a decrease in \ref{p3} (\greenmark). However, \ref{obj2-subgroups} is likely to be the dominant factor. Hence, the nature of this interaction will be determined by whether the distinguishability in observables across subgroups increases or decreases with \ref{rd1}.

\noindent\textbf{\ref{rd2} (Outlier Removal) and \ref{p2} (Data Reconstruction).}
Here, \ref{data1} is the common factor and the arrows are in the same direction, suggesting an increase in \ref{p2} (\redmark). Similar to membership and attribute inference~\cite{duddu2021shapr,carlini2022the,jayaraman2022attribute}, we expect retraining without outliers will make $\noOutlierModel$ memorize a new set of data records, thus making them more susceptible to \ref{p2}.

\noindent\textbf{\ref{rd2} (Outlier Removal) and \ref{p4} (Distribution Inference).} \ref{data2} is the common factor which has arrows in the same direction which suggests an increase in \ref{p4} (\redmark). Prior work has shown that adding more outliers is likely to increase \ref{p4}~\cite{chaudhari2022snap,chase2021property}. \ref{rd2} removes outliers belonging from tail classes. We can assume that these outliers belong to minority subgroup, and outlier removal will increase the bias thereby increasing the distinguishability in observables for datasets with different properties. Hence, \ref{p4} is likely to increase.

%% file: tables/tab_unexplored.tex
\begin{table}[!htb]
\caption{Unexplored interactions (\graymark\xspace from Table~\ref{tab:interactions}) from a defense to a risk (indicated with  ``\ding{222}'' between them).}
\begin{center}
\footnotesize
\begin{tabular}{l}
\hline
\textbf{Defense} \ding{222} \textbf{Risk}\\
\hline
 \ref{rd1} (Adversarial Training) \ding{222} \ref{p3} (Attribute Inference)  \\ 
\ref{rd2} (Outlier Removal) \ding{222} \ref{r3} (Unauthorized Model Ownership)\\
\ref{rd2} (Outlier Removal) \ding{222} \ref{p2} (Data Reconstruction)\\
\ref{rd2} (Outlier Removal) \ding{222} \ref{p4} (Distribution Inference)\\
\ref{rd3} (Watermarking) \ding{222} \ref{r1} (Evasion)\\ 
\ref{rd4} (Fingerprinting) \ding{222} \ref{r2} (Poisoning) \\
\ref{rd4} (Fingerprinting) \ding{222} \ref{p2} (Data Reconstruction)\\
\ref{rd4} (Fingerprinting) \ding{222} \ref{p3} (Attribute Inference)\\ 
\ref{rd4} (Fingerprinting) \ding{222} \ref{p4} (Distribution Inference)\\
\ref{rd4} (Fingerprinting) \ding{222} \ref{u} (Discriminatory Behaviour)\\

\ref{pd1} (Differential Privacy) \ding{222} \ref{r3} (Unauthorized Model Ownership)\\ 
\ref{ud1} (Group Fairness) \ding{222}  \ref{r3} (Unauthorized Model Ownership)\\
\ref{ud1} (Group Fairness) \ding{222} \ref{p2} (Data Reconstruction)\\ 
\ref{ud2} (Explanations) \ding{222} \ref{p4} (Distribution Inference)\\
\hline
\end{tabular}
\end{center}
\label{tab:unexplored}
\end{table}

%% file: 6related.tex
\section{Related Work}\label{sec:related}

\noindent\textbf{Surveys/SoKs.} There are several surveys covering defenses and risks separately~\cite{SoKMLPrivSec,robustSoK,certifiedRobustness2020sok,privacySurvey,miaSurvey,salem2022sok,fairSurvey,transparencySurvey}. None of them explore the interactions between defenses and risks. Some prior works cover interactions among defenses and risks. Strobel and Shokri~\cite{shokriSurvey} illustrate the limited case of increased membership inference with robustness, fairness, and explanations. Noppel and Wressnegger~\cite{sokExpl} cover specific interaction between explanations and evasions/poisoning. Chen et al.~\cite{fedtradeoff} present conflicts between fairness and different privacy risks. Gittens et al.~\cite{surveyTradeoff} study the interactions of different risks with DP, fairness and robustness but do not cover the factors underlying these interactions. Ferry et al.~\cite{ferry:sok} evaluate the interactions among fairness, privacy and interpretability but do not cover the factors underlying these interactions.

\noindent\textbf{Interactions within Defenses.} An orthogonal line of work evaluates trade-offs within defenses or designs algorithms to improve them.
For instance, it is impossible to achieve group fairness and DP together~\cite{cummingsDPFair,alves2023survey,fioretto2022differential}. Some prior works attempt to reconcile them by offering pareto-optimal solutions with an accuracy drop~\cite{dpfair,yaghini2023learning}. Further, privacy and transparency are inherently conflicting and designing explanations with DP degrades the quality of explanations~\cite{DPExplanations,dpexplanations2,dpexplanations3}. Conflicts also arise between watermarking with adversarial training and DP~\cite{szyller2022conflicting}.
Gradients from models with adversarial training are interpretable which suggest potential alignment between transparency and robustness~\cite{robustVacc}. 
Furthermore, prior works have designed objective functions to reconcile adversarial training and group fairness~\cite{robustOrFair}.
Finally, DP over image pixels can provide certified robustness against evasion~\cite{lecuyerCertified}. 

%% file: 7discussion.tex
\section{Discussion and Conclusions}\label{sec:discussion}

\noindent\textbf{Completeness of the framework.} We identified overfitting and memorization (and their underlying factors) as the causes for unintended interactions based on prior work. These are sufficient to explain all the unintended interactions explored in the literature as seen in our survey (Sections~\ref{sec:framework}--\ref{sec:interactions}). We do not claim them to be complete considering the complexity of ML models. 
Whether there are indeed other possible underlying causes, remains an open question.
Regardless, our framework is flexible enough to accommodate emerging causes and factors. To incorporate a new cause, we can use the same methodology as in Section~\ref{sec:framework}--\ref{sec:interactions}: identify factors that influence the cause, describe how they relate with defenses and risks, and use them to conjecture unintended interactions.
Further, the framework can be extended to cover variants of current risks and defenses as well as new ones by expanding Tables~\ref{tab:factorRelation} and~\ref{tab:interactions}. 
Finally, the framework can be extended to account for any exceptions in some interactions stemming from a difference in threat model.

\noindent\textbf{Interactions not covered by our Guideline.} There are a few interactions which are not covered by our guideline which can be addressed by extending our framework:
\begin{itemize}[leftmargin=*]
\item \ref{rd2} (Outlier Removal), \ref{r2} (Poisoning), and \ref{r3} (Unauthorized Model Ownership) have too few (non-dominant) factors to confidently identify the nature of some interactions:
\ref{rd1} (Adversarial training) \ding{222} \ref{r2} (Poisoning) and \ref{r3} (Unauthorized Model Ownership); \ref{rd2} (Outlier Removal) \ding{222} \ref{r1} (Evasion) and \ref{r2} (Poisoning); \ref{pd1} (Differential Privacy) \ding{222} \ref{r2} (Poisoning); \ref{ud1} (Group fairness) \ding{222} \ref{r2} (Poisoning); and \ref{ud2} (Explanations) \ding{222} \ref{r2} (Poisoning), \ref{r3} (Unauthorized Model Ownership). Additionally, \ref{ud2} (Explanations) \ding{222} \ref{r1} (Evasion) do not have any common factors to determine the interaction using our guideline.
\item \ref{rd1} (Adversarial Training) and \ref{pd1} (Differential Privacy) \ding{222} \ref{u} (Discriminatory Behaviour): \ref{data2} is a common factor but it covers distribution over different classes but it does not account for subgroups. That is, data records from the tail classes may not belong to a minority subgroup. 
% \item \ref{rd2} (Outlier Removal) \ding{222} \ref{r1} (Evasion), \ref{r2} (Poisoning): \ref{data2} suggests an increase in risks (\redmark) but empirical evaluation suggests a decrease in risks (\greenmark): \ref{rd2} combines both removing outliers from $\dtrain$ and retraining. 
% However, for \ref{r1} and \ref{r2} do not require retraining.
\end{itemize}

\noindent\textbf{Minimizing unintended interactions.} Given the factors influencing unintended interactions, practitioners can mitigate unintended interactions by exploring:
\begin{enumerate*}[leftmargin=*,itemjoin={,\xspace}]
\item data augmentation (\ref{data1})
\item balanced classes with undersampling and oversampling (\ref{data2})
\item adding proxy attributes or removing redundant attributes (\ref{data3})
\item domain generalization algorithms (e.g., invariant risk minimization)~\cite{mahajan2021connection,hartmann2022distribution} (\ref{data4})
\item adjusting training hyperparameters (\ref{obj1}, \ref{obj2}, and \ref{obj3} )
\item experimenting with more layers, lower precision, or pruning (\ref{model1}).
\end{enumerate*}

\noindent\textbf{Applicability beyond classifiers.} Our framework extends to other models by:
\begin{enumerate*}[leftmargin=*,itemjoin={,\xspace}]
\item identifying unintended interactions
\item evaluating the relation between defenses, risks, and underlying factors
\item including them in our framework for insights using our guideline.
\end{enumerate*}
For example, in generative models where overfitting and memorization manifest as the replication of $\dtrain$~\cite{secretSharer,extractDiffusion,extractlm}, \ref{data1} and \ref{model1} are relevant.

\noindent\textbf{Future work.} 
In on-going work, we are creating a software framework to facilitate systematic empirical exploration of unintended interactions.
In addition to empirically evaluating different unexplored interactions (Table~\ref{tab:unexplored}), a more comprehensive evaluation of how factors correlate with effectiveness of defenses and susceptibility to risks is needed (i.e., expanding Table~\ref{tab:factorRelation}). Similarly, understanding how different variants of risks and defenses (e.g., types of fairness algorithms and metrics) affect the interaction, needs further study. Also, our framework suggests correlation and not causation. \emph{Causal reasoning} can establish causation between different factors with unintended interactions (e.g.,~\cite{baluta2022membership}). Finally, the interactions are empirically evaluated, both in existing literature and our experiments in Section~\ref{sec:conjectures}. Formally modeling these interactions and predicting their nature mathematically is for future work.

\noindent\textbf{Takeaways.} We highlight the following key takeaways:
\begin{enumerate*}[label=\roman*),itemjoin={,\xspace}]
\item Current literature lacks a comprehensive understanding of unintended interactions between defenses (intended to protect against some risks), and \emph{other unrelated risks}. But understanding such interactions is critical for deployment. We showed that identifying common factors (for overfitting and memorization) between a defense and a risk can shed light on whether these factors can induce unintended interactions
% \item a software framework can aid in the systematic empirical exploration of unintended interactions 
\item an important open problem is: how to design effective defenses and select the right combination of defenses to minimize increases in susceptibility to other risks?
\end{enumerate*}

%% file: 8appendix.tex
\appendices

\section{Summary of Notations}\label{sec:notations}

\input{tables/tab_notations}

\newpage
\section{Meta-Review}

\subsection{Summary of Paper}

The paper study a broad range of risks and their (unintended) induction by defenses. Specifically, the authors observe that the risks associated with machine learning are vast, and while there have been many proposed defenses to address their shortcomings, papers have (informally) observed that defenses may induce other deficiencies explored in other threat models. To this end, the authors propose a systematic framework that categories risks across security and privacy, while simultaneously linking them where empirical or theoretical evidence has been found. With this broad perspective, the authors identify many opportunities for future research, including the lack of evidence for some interactions as well as some interactions that have yet to be measured. Using their framework, the authors perform a set of experiments to validate some conjectures made apparent by the framework and validate their hypotheses empirically. Finally, the authors also provide general guidelines for how unintended interactions can be identified.

\subsection{Scientific Contributions}
\begin{itemize}[leftmargin=*]
    \item 1. Independent Confirmation of Important Results with Limited Prior Research
    \item 6. Provides a Valuable Step Forward in an Established Field
    \item 7. Establishes a New Research Direction
\end{itemize}

\subsection{Reasons for Acceptance}
\begin{enumerate}[leftmargin=*]
\item The core scientific contribution of this work is the framework that unifies vast bodies of work across machine learning security and privacy risks. The framework offers convincing evidence on why interactions happen, what has left to be identified, and how future work can systematically explore if a defense induces an unintended interaction.
\item The paper did an excellent job surveying, collecting, and extrapolating the interactions between ML defenses. The bibliography is comprehensive. This paper can greatly impact the implementation of ML defenses and might open new research possibilities for the exploitation of these interactions.
\item The presentation of this paper is exceptional. The ideas are concise, the definitions are rooted in well-grounded concepts of learning theory as well as modern definitions commonly observed in security and privacy literature.
\end{enumerate}

%% file: tables/tab_notations.tex
% \begin{arxiv}
\begin{table}[!htb]
\caption{Summary of notations and their descriptions.}
\begin{center}
\tiny
\begin{tabular}{ |c|c| } 
 \hline
 \textbf{Notation} & \textbf{Description}\\
  \hline
 \multicolumn{2}{|c|}{\textbf{Standard Model and Metrics}}\\
 \hline
 $\model$ & standard ML model (no defense)  \\ 
 $\theta$ & $\model$'s parameters\\
  $\mathcal{A}$ & Training algorithm\\
 $R(\theta)$ & Regularization function over $\theta$\\
  $\lambda$ & Regularization hyperparameter\\
   $l(\model,y)$ & Loss function\\
 $f^k_{\theta}(x)$ & $k^{th}$ layer's output (activation)\\
 $\mathcal{C}$ & Cost function to be optimized during training\\
 $\attmodel$ & Adversary's attack model\\
 $\mathcal{G}_{err}$ & Generalization error (train-test accuracy gap)\\
 \hline
 \multicolumn{2}{|c|}{\textbf{Data}}\\
 \hline
 $\dtrain$ & Training dataset for $\model$  \\
 $z_i=(x_i,y_i)$ & Data record with attributes $x_i$ and label $y_i$\\
 $\mathcal{S}$ & Random subset of $\dtrain$\\
 $|\dtrain|$ & Size of $\dtrain$ \\
 $\dtest$ & Test dataset  \\ 
 $\dtrainadv$ & Adversary's dataset to train $\attmodel$ \\
 $\dtestadv$ & Adversary's dataset to evaluate $\attmodel$ \\
 \hline
  \multicolumn{2}{|c|}{\textbf{Defenses and Risks}}\\
 \hline
$\robustmodel$ & Robust model from adversarial training  \\ 
$\privatemodel$ & Private model from DPSGD \\
$\fairmodel$ & Fair model with group fairness constrains \\
 $\derivedmodel$ & Stolen model derived from $\model$ \\
 $x_{adv}$ & Adversarial Example\\
  $\noOutlierModel$ & Model after retraining on $\dtrain$ without outliers\\
 $\watermarkModel$ & Model trained with watermarks\\
 $\epsilon_{rob}$ & Perturbation budget for adversarial examples\\
 $\delta_{rob}$ & Perturbation added to generate $x_{adv}$\\
 $\epsilon_{dp}$ & Privacy budget\\
 $\delta_{dp}$ & probability where privacy loss $>$ $e^\epsilon_{dp}$\\
\hline
\multicolumn{2}{|c|}{\textbf{Evaluating Conjectures}}\\
\hline
 
 $\alpha_1$,  $\alpha_2$ & Distributional properties of $\dtrain$\\
 $^{\alpha_1}\dtrainadv$, $^{\alpha_2}\dtrainadv$ & Datasets satisfying $\alpha_1$,  $\alpha_2$\\
 $\phi(x)^{\alpha_1}$, $\phi(x)^{\alpha_2}$ & Explanations corresponding to  $\alpha_1$ and $\alpha_2$\\
 \hline
\end{tabular}
\end{center}
\label{tab:notations}
\end{table}
% \end{arxiv}

% \begin{submit}
% \begin{table}[!htb]
% \caption{Frequently used notations and their descriptions.}
% \begin{center}
% \footnotesize
% \begin{tabular}{ |c|c| } 
%  \hline
%  \textbf{Notation} & \textbf{Description}\\
%  \hline
%   $\dtrain$ & Training dataset for $\model$  \\
%  $\mathcal{G}_{err}$ & Generalization error (train-test accuracy gap)\\
%  $\model$ & standard ML model (no defense)  \\ 
%  $\theta$ & $\model$'s parameters\\
% $\robustmodel$ & Robust model from adversarial training  \\ 
% $\privatemodel$ & Private model from DPSGD \\
% $\fairmodel$ & Fair model with group fairness constrains \\
%  $\derivedmodel$ & Stolen model derived from $\model$ \\
% $\noOutlierModel$ & Model after retraining without outliers\\
%  $\watermarkModel$ & Model trained with watermarks\\
%   $\epsilon_{rob}$ & Perturbation budget for adversarial examples\\
%  $\epsilon_{dp}$ & Privacy budget for DP\\
% \hline
% \end{tabular}
% \end{center}
% \label{tab:notations}
% \end{table}
% \end{submit}